\documentclass[]{pasj00}

\newcommand{\kms}{\mbox{km s$^{-1}$}}

\draft
\begin{document}
\SetRunningHead{Y. Mizuno et al.}{$^{12}$CO $J$=4--3 observation towards the LMC-N159}
\Received{2009/1/5}

\title{Warm and Dense Molecular Gas in the N159 Region:\\
$^{12}$CO $J$=4--3 and $^{13}$CO $J$=3--2 Observations\\
 with NANTEN2 and ASTE}
\author{Yoji \textsc{Mizuno}\altaffilmark{1},
Akiko \textsc{Kawamura}\altaffilmark{1},
Toshikazu \textsc{Onishi}\altaffilmark{1,2},
Tetsuhiro \textsc{Minamidani}\altaffilmark{1,3},
Erik \textsc{Muller}\altaffilmark{1},
Hiroaki \textsc{Yamamoto}\altaffilmark{1},
Takahiro \textsc{Hayakawa}\altaffilmark{1},
Norikazu \textsc{Mizuno}\altaffilmark{1,4},
Akira \textsc{Mizuno}\altaffilmark{5},
J\"{u}rgen \textsc{Stutzki}\altaffilmark{6},
Jorge L. \textsc{Pineda}\altaffilmark{7},
Uli \textsc{Klein}\altaffilmark{8},
Frank \textsc{Bertoldi}\altaffilmark{8},
Bon-Chul \textsc{Koo}\altaffilmark{9},
Monica \textsc{Rubio}\altaffilmark{10},
Michael \textsc{Burton}\altaffilmark{11},
Arnold \textsc{Benz}\altaffilmark{12},
Hajime \textsc{Ezawa}\altaffilmark{4},
Nobuyuki \textsc{Yamaguchi}\altaffilmark{4},
Kotaro \textsc{Kohno}\altaffilmark{13},
Tetsuo \textsc{Hasegawa}\altaffilmark{4},
Ken'ichi \textsc{Tatematsu}\altaffilmark{4},
Masafumi \textsc{Ikeda}\altaffilmark{4},
J\"{u}rgen \textsc{Ott}\altaffilmark{14,15},
Tony \textsc{Wong}\altaffilmark{16},
Annie \textsc{Hughes}\altaffilmark{17,18},
Margaret \textsc{Meixner}\altaffilmark{19},
Remy \textsc{Indebetouw}\altaffilmark{20,21},
Karl D. \textsc{Gordon}\altaffilmark{19},
Barbara \textsc{Whitney}\altaffilmark{22},
Jean-Philippe \textsc{Bernard}\altaffilmark{23},
and 
Yasuo \textsc{Fukui}\altaffilmark{1}
}

\altaffiltext{1}{Department of Astrophysics, Nagoya University, Furocho, Chikusaku, Nagoya 464-8602, Japan}
\altaffiltext{2}{Department of Physical Science, Osaka Prefecture University, Gakuen 1-1, Sakai, Osaka 599-8531, Japan}
\altaffiltext{3}{Department of Physics, Faculty of Science, Hokkaido University, N10W8, Kita-ku, Sapporo 060-0810, Japan.}
\altaffiltext{4}{National Astronomical Observatory of Japan, Mitaka, Tokyo 181-8588, Japan}
\altaffiltext{5}{Solar-terrestrial Environment Laboratory, Nagoya University, Furocho, Chikusaku, Nagoya 464-8601, Japan}
\altaffiltext{6}{KOSMA, I. Physikalisches Institue, Universit\"{a}t zu K\"{o}ln, Z\"{u}lpicher Stra$\beta$e 77, 50937 K\"{o}ln, Germany.}
\altaffiltext{7}{Jet Propulsion Laboratory, M/S 169-507, 4800 Oak Grove Drive, Pasadena, CA 91109., USA.}
\altaffiltext{8}{Argelander-Institut f\"{u}r Astronomie, Auf dem H\"{u}gel 71, 53121 Bonn, Germany.}
\altaffiltext{9}{Seoul National University, Seoul 151-742, Korea.}
\altaffiltext{10}{Departamento de Astronomia, Universidad de Chile, Casilla 36-D, Santiago, Chile.}
\altaffiltext{11}{School of Physics, University of New South Wales, Sydney 2052, NSW, Australia.}
\altaffiltext{12}{Institute of Astronomy, ETH Zurich, 8093 Zurich, Switzerland.}
\altaffiltext{13}{Institute of Astronomy, University of Tokyo, Mitaka, Tokyo 181-0015, Japan}
\altaffiltext{14}{National Radio Astronomy Observatory, P.O. Box O, Socorro, NM, USA}
\altaffiltext{15}{California Institute of Technology, 1200 E. California Blvd., Caltech Astronomy, 105-24 Pasadena, CA 91125-2400, USA.}
\altaffiltext{16}{Department of Astronomy, University of Illinois, 1002 W. Green St., Urbana, IL 61801, USA.}
\altaffiltext{17}{CSIRO Australia Telescope National Facility, P.O. Box 76, Epping, NSW 1710, Australia.}
\altaffiltext{18}{Center for Supercomputing and Astrophysics, Swinburne University of Technology, Hawthorn, VIC 3122, Australia.}
\altaffiltext{19}{Space Telescope Science Institute, 3700 San Martin Way, Baltimore, MD 21218, USA}
\altaffiltext{20}{Department of Astronomy, University of Virginia, P.O. Box 400325, Charlottesville, VA 22903, USA}
\altaffiltext{21}{National Radio Astronomy Observatory, 520 Edgemont Rd., Charlottesville, VA 22903, USA}
\altaffiltext{22}{Space Science Institute, 4750 Walnut St. Suite 205, Boulder, CO 80301, USA}
\altaffiltext{23}{CESR,9 Av du Colonel Roche, BP 44346, 31028 Toulouse cedex 4, France}
\email{y\_mizuno@a.phys.nagoya-u.ac.jp}

\KeyWords{Magellanic Clouds --- ISM: clouds --- ISM: molecules --- radio lines: ISM --- submillimeter --- stars: formation}
\maketitle

\begin{abstract}
New $^{12}$CO $J$=4--3 and $^{13}$CO $J$=3--2 observations of the N159 region, an active site of massive star formation in the Large Magellanic Cloud, have been made with the NANTEN2 and ASTE sub-mm telescopes, respectively. The $^{12}$CO $J$=4--3  distribution is separated into three clumps, each associated with N159W, N159E and N159S. These new measurements toward the three clumps are used in coupled calculations of molecular rotational excitation and line radiation transfer, along with other transitions of the $^{12}$CO $J$=1--0, $J$=2--1, $J$=3--2, and $J$=7--6 as well as the isotope transitions of $^{13}$CO $J$=1--0, $J$=2--1, $J$=3--2, and $J$=4--3. The $^{13}$CO $J$=3--2 data are newly taken for the present work. The temperatures and densities are determined to be $\sim$ 70-80K and $\sim 3 \times 10^3$ cm$^{-3}$ in N159W and N159E and $\sim$ 30K and $\sim 1.6 \times 10^3$ cm$^{-3}$ in N159S. Observed $^{12}$CO $J$=2--1 and $^{12}$CO $J$=1--0 intensities toward N159W and N159E are weaker than expected from calculations of uniform temperature and density, suggesting that low-excitation foreground gas causes self-absorption. 
These results are compared with the star formation activity based on the data of young stellar clusters and H\emissiontype{II} regions as well as the mid-infrared emission obtained with the Spitzer MIPS. The N159E clump is associated with embedded cluster(s) as observed at 24 $\micron$ by the Spitzer MIPS and the derived high temperature, 80K, is explained as due to the heating by these sources.  The N159E clump is likely responsible for a dark lane in a large H\emissiontype{II} region by the dust extinction. On the other hand, the N159W clump is associated with embedded clusters mainly toward the eastern edge of the clump only. These clusters show offsets of 20$\arcsec$ - 40$\arcsec$ from the $^{12}$CO $J$=4--3 peak and are probably responsible for heating indicated by the derived high temperature, 70 K. The N159W clump exhibits no sign of star formation toward the $^{12}$CO $J$=4--3 peak position and its western region that shows enhanced $R_{4-3/1-0}$ and $R_{3-2/1-0}$ ratios. We therefore suggest that the N159W peak represents a pre-star-cluster core of $\sim 10^5 M_{\odot}$ which deserves further detailed studies. We note that recent star formation took place between N159W and N159E as indicated by several star clusters and H\emissiontype{II} regions, while the natal molecular gas toward the stars have already been dissipated by the ionization and stellar winds of the OB stars. The N159S clump shows little sign of star formation as is consistent with the lower temperature, 30K, and somewhat lower density than N159W and N159E. The N159S clump is also a candidate for future star formation.
\end{abstract}

\section{Introduction}
Giant molecular clouds (hereafter GMCs) are the principal sites of star formation and studies of GMCs are important in understanding the evolution of galaxies. A few tens of GMCs in the solar neighborhood such as Orion A have been well studied, while most of the GMCs in the Galaxy are located at more than a few kpc away in the Galactic disk, where contamination by un-related components in the same line of sight seriously limits detailed understanding of GMCs and their associated objects.

Recent CO surveys of molecular clouds toward external galaxies in the Local Group have revealed that properties of GMCs such as relations between CO luminosity and line width, the mass range, and the index of mass spectrum are similar among these galaxies \citep{2007prpl.conf...81B}. This suggests that the properties of GMCs are fairly common among galaxies. The Large Magellanic Cloud (hereafter LMC) is the nearest galaxy to our own at a distance of 50 kpc and is nearly face on, making it an ideal laboratory to observe various properties of GMCs. The small distance enables us to make molecular observations at high spatial resolutions. The LMC also provides a unique opportunity to study molecular clouds and star formation in different environments from the Galaxy; the gas-to-dust ratio is $\sim 4$ times higher \citep{1982A&A...107..247K}, and the metal abundance is about $\sim $3 -- 4 times lower \citep{2002A&A...396...53R,1984IAUS..108..353D} than those of the Galaxy. 

Following a low resolution $^{12}$CO $J$=1--0 survey at 140 pc resolution with the Columbia 1.2 m telescope located at CTIO in Chile \citep{1988ApJ...331L..95C}, high resolution observations in the $^{12}$CO $J$=1--0, $J$=2--1 emission line were made with the SEST 15 m telescope toward some of the molecular clouds and revealed their clumpy structure at 10 pc resolution (e.g., \cite{1986ApJ...303..186I,1994A&A...291...89J,1996ApJ...472..611C,1997A&AS..122..255K,1998A&A...331..857J,2003A&A...406..817I}). However, these high resolution studies were limited in the spatial coverage, compared with the large angular extent of the LMC, $\sim 6\arcdeg \times 6\arcdeg$. \citet{2008ApJS..178...56F} carried out a survey in the $^{12}$CO $J$=1--0 emission line at 40 pc resolution over a $6\arcdeg \times 6\arcdeg$ field in the LMC with the NANTEN 4 m telescope and obtained a complete sample of 270 GMCs (see also for preceding works \cite{1999PASJ...51..745F,2001PASJ...53L..41F,2001PASJ...53..971M,2001PASJ...53..985Y}). These studies revealed that young stellar clusters whose ages are less than 10 Myr are spatially well correlated with GMCs, and that GMCs are categorized into three types in terms of star formation activity. Type I is starless in the sense that they are not associated with O stars, Type II is associated with small H\emissiontype{II} regions only, and Type III is associated with huge H\emissiontype{II} regions and young stellar clusters \citep{Kawamura09, 2007prpl.conf...81B}. These types are interpreted in terms of evolutionary sequence of GMCs in a timescale of 2-30 Myrs \citep{1999PASJ...51..745F,2007prpl.conf...81B,Kawamura09}.

The $^{12}$CO $J$=1--0 emission line is a probe commonly used to trace molecular clouds because of its low excitation energy ($\sim$ 5K) and low critical density for collisional excitation ($n_{cr} \sim 1000$cm$^{-3}$). The $^{12}$CO $J$=1--0 emission alone is however not able to provide physical properties like kinetic temperature and density, the fundamental parameters of GMCs. The high $J$ transitions have higher excitation energies and higher critical densities; e.g., the $^{12}$CO $J$=3--2 transition has the upper state at 33 K and the critical density, $3\times 10^4 $cm$^{-3}$ \citep{2005A&A...432..369S}, and the $^{12}$CO $J$=4--3 transition has the upper state at 55 K and the critical density, $1\times 10^5 $cm$^{-3}$. These sub--mm CO emission lines can selectively trace the sites which may be warmer and denser than the mm CO lines and have a potential to reveal detailed physical properties where star formation is taking place. We are also allowed to make better estimates of temperatures and densities of molecular clouds with a combination of multi--$J$ CO line intensities and molecular excitation analyses such as the Large Velocity Gradient (hereafter LVG) model of molecular line transfer. Recently-developed sub--mm telescopes such as ASTE, NANTEN2 and APEX located at altitudes around 5000m in Atacama, Chile have enabled us to observe higher excited CO transitions at sub--mm wavelengths in superb observational conditions \citep{2004SPIE.5489..763E,2006IAUSS...1E..21F,2006A&A...454L.115G}. At a lower angular resolution, AST/RO also offered sub-mm observing capability \citep{2001PASP..113..567S}.

 \citet{2000ApJ...545..234B} first detected the $^{12}$CO $J$=4--3 emission line toward the N159 region at 50 pc resolution with the AST/RO telescope. \citet{2005ApJ...633..210B} later presented estimates of temperatures and densities; $T_{\rm kin}$ = 20 K, $n({\rm H}_2)$ = 10$^5$ cm$^{-3}$ for the cold dense component and $T_{\rm kin}$ = 100 K, $n({\rm H}_2)$ = 100 cm$^{-3}$ for hot tenuous component toward N159W. Subsequently, \citet{2008ApJS..175..485M} carried out high resolution $^{12}$CO $J$=3--2 observations of several GMCs at 5 pc resolution with the ASTE 10 m telescope including N159 in the LMC. \citet{2008ApJS..175..485M} revealed detailed structure of highly excited gas with $T_{\rm kin} \sim$ 20 -- 200 K and $n({\rm H}_2)$ $\sim$ 10$^3$ -- 10$^4$ cm$^{-3}$.
 In the present study, we shall focus on the N159 GMC which shows the highest $^{12}$CO $J$=1--0 intensity from the NANTEN survey \citep{2008ApJS..178...56F}. The N159 GMC is classified as Type III, and includes at least three prominent clumps, N159W, N159E and N159S \citep{1994A&A...291...89J}. The two molecular clumps in the northern part, N159W and N159E, are associated with massive star clusters whose ages are younger than 10 Myr \citep{1996ApJS..102...57B} and with huge H\emissiontype{II} regions \citep{1956ApJS....2..315H,1976MmRAS..81...89D,1986ApJ...306..130K}. On the other hand, there is no star formation in N159S. In order to better estimate the physical properties of the N159 region, we have carried out new sub-mm observations in the $^{12}$CO $J$=4--3 emission line at a 10 pc spatial resolution with NANTEN2 and in the $^{13}$CO $J$=3--2 emission line at a 5 pc spatial resolution with ASTE. We have also made combined calculations of molecular rotational excitation and line radiative transfer by employing the CO datasets available in N159 and compared the results with the star formation activity. 

The present paper is organized as follows: Section \ref{pasj_obs} describes the observations. Sections \ref{pasj_res} and \ref{pasj_analysis} show the observational results and data analysis, respectively. We discuss the correlation between highly excited molecular gas and star formation activities in section \ref{pasj_sf}. Finally, we present a summary in section \ref{pasj_sum}.

\section{Observations}\label{pasj_obs}
 \subsection{$^{12}$CO $J$=4--3 observations}
We observed the $^{12}$CO $J$=4--3 rotational line at 461.0408 GHz with the NANTEN2 \citep{2006IAUSS...1E..21F} 4 m sub-mm telescope situated at a 4800 m altitude at Pampa la Bola in Chile. The telescope has a main beam efficiency of 0.6 and the half power beam width (HPBW) of 38{\arcsec}, as determined from continuum cross scans on Jupiter at 461 GHz. The forward beam coupling efficiency was measured by sky-dip measurements to be 0.86. We used a dual-channel 460 / 810 GHz receiver developed by University of Cologne. Observations were made during September to December in 2006 for N159W and October to December in 2007 for the N159E and N159S regions. The double sideband receiver temperature was measured to be 250 K at 460 GHz and the typical system noise temperature during these observations was $\sim$ 900 K toward the zenith. The spectrometer was an acousto--optical spectrometer (AOS) with total bandwidth of 1 GHz divided into 1730 channels, yielding a velocity resolution of 0.37 \kms. We observed three 2\arcmin $\times$2{\arcmin} areas toward the three prominent clumps; N159W, N159E and N159S \citep{1998A&A...331..857J}. Observations were made using  on--the--fly (OTF) mode with two orthogonal scan directions along RA and DEC  to ensure a flat field and the data were gridded with a 10{\arcsec} spacing. Calibration and data gridding were made with the ``{\sc kalibrate}'' software developed by University of Cologne. The pointing accuracy was better than 10{\arcsec}, as confirmed from pointing measurements on IRC2 in Orion A. The final noise levels are 1.2 K/ch for N159W and 0.4 K/ch for N159E and N159S.

 \subsection{$^{13}$CO $J$=3--2 Observations}
The $^{13}$CO $J$=3--2 rotational line at 330.5879 GHz was observed with the ASTE 10 m sub-mm telescope in Pampa la Bola in Chile \citep{2004SPIE.5489..763E,2008SPIE.7012E...6E}, using a double side band superconductor insulator superconductor (SIS) mixer receiver \citep{2007PASJ...59...43M} in a period from 20 to 23 September 2006. The half--power beam width was measured from observations of Jupiter to be 23{\arcsec}. The spectrometer was an XF--type digital auto correlator \citep{2000SPIE.4015...86S} and was operated in the wideband mode, having a bandwidth of 512 MHz with 1024 channels. This configuration yielded a velocity coverage and resolution of 450 and 0.45 \kms, respectively. We observed 3 $\times$ 3 positions at 20{\arcsec} spacing in the position switching mode centered at the $^{12}$CO $J$=3--2 integrated intensity peaks \citep{2008ApJS..175..485M} in N159W ($5^{\rm h} 39^{\rm m} 36.8^{\rm s}$, -69{\arcdeg} 45{\arcmin} 32{\arcsec} at J2000 coordinate), N159E ($5^{\rm h} 40^{\rm m} 8.7^{\rm s}$, -69{\arcdeg} 44{\arcmin} 34{\arcsec}) and N159S ($5^{\rm h} 40^{\rm m} 5^{\rm s}$, -69{\arcdeg} 50{\arcmin} 34{\arcsec}), where the observing grid is along the B1950 coordinate. The pointing error was measured to be within 5{\arcsec} from the observations of the $^{12}$CO $J$=3--2 point source R--Dor. Typical system noise temperature was measured to be 260 K in DSB. Finally, the map was re--gridded to the J2000 coordinate by using ``{\sc miriad}'' astronomical data analyzing software. We observed Ori--KL and M17SW to check the stability of the intensity calibration, and the maximum intensity variation during these observations was 12\%. We use 0.7 for main beam efficiency, as measured from observations of Jupiter. Due to a lack of side--band ratio data, the calibrated intensity scale was somewhat uncertain in 2006, and we re-observed the peak position of N159W with the newly implemented single side band receiver \citep{2008SPIE.7012E...6E} in 2008, and estimated the main beam intensity to be 3.15 $\pm$ 0.15 (K). We scaled the observational data in 2006 by using the derived scaling factor. Achieved noise levels in the main beam temperature scale are 0.5 K/ch for N159W and 0.2 K/ch for N159E and N159S.

\section{Observational Results}\label{pasj_res}
\subsection{The $^{12}$CO $J$=4--3 and $^{13}$CO $J$=3--2 emission lines}\label{chap3_res1}

\begin{figure}
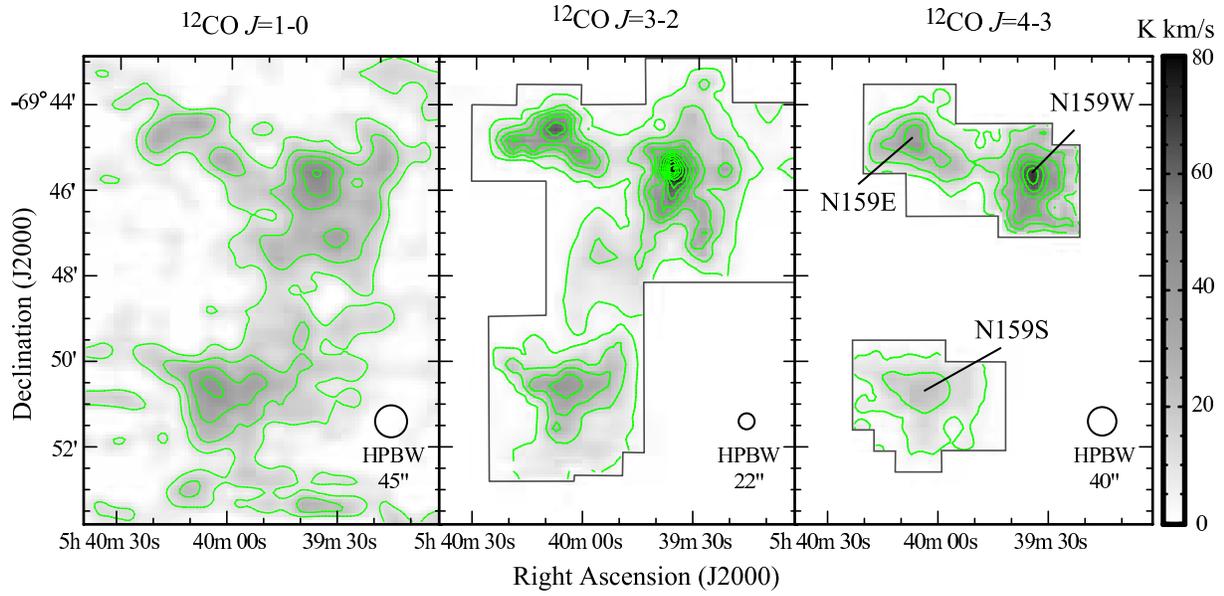

\begin{center}
\FigureFile(160mm,80mm){figure1.eps}
\caption{Integrated intensity map of $^{12}$CO $J$=1--0 (left), $^{12}$CO $J$=3--2 (middle) and $^{12}$CO $J$=4--3 (right). Lowest contour levels are 3$\sigma$ (11, 5, 5 K \kms for $^{12}$CO $J$=1--0, $^{12}$CO $J$=3--2 and $^{12}$CO $J$=4--3 respectively) level of each observation and increments are 10 K \kms.}
\label{fig:1}
\end{center}
\end{figure}

Figure \ref{fig:1} shows the spatial distributions of the $^{12}$CO emission lines. The right panel of Figure \ref{fig:1} shows the distribution of the $^{12}$CO $J$=4--3 integrated intensity smoothed to a 40{\arcsec} beam, assuming a Gaussian beam shape. For comparison, the left and middle panels of Figure \ref{fig:1} show the distributions of the total integrated intensity of the $^{12}$CO $J$=1--0 \citep{2008PASA...25..129O} and the $^{12}$CO $J$=3--2 \citep{2008ApJS..175..485M}, respectively. Distributions of these $^{12}$CO emission are  largely similar with each other. The $^{12}$CO $J$=4--3 distribution has clearly resolved the cloud into three clumps, N159W, N159E, and N159S which were identified in the $^{12}$CO $J$=1--0 observations \citep{1994A&A...291...89J}. These three were not resolved in the previous $^{12}$CO $J$=4--3 observations \citep{2000ApJ...545..234B}.

 The $^{13}$CO $J$=3--2 emission was measured in a 3 $\times$ 3 pointing pattern, with a 20{\arcsec} grid spacing toward each $^{12}$CO $J$=3--2 peak of N159W, N159E and N159S. The line profiles are shown in Figure \ref{fig:2}. 

Figure \ref{fig:3} shows four spectra of the $^{12}$CO $J$=1--0, $J$=3--2 and $J$=4--3 and $^{13}$CO $J$=3--2 emission lines toward three peaks of the $^{12}$CO $J$=3--2 integrated intensity \citep{2008ApJS..175..485M}. The spectra are convolved to a 45{\arcsec} Gaussian beam for a quantitative comparison. N159W shows the strongest emission among the four spectra. 

Table \ref{tbl:1} lists the observed parameters of the $^{12}$CO $J$=4--3 and $^{13}$CO $J$=3--2 emission lines toward selected peak positions of the N159 region. The parameters of the other emission lines including $^{12}$CO $J$=1--0, $J$=2--1, $J$=3--2 and $J$=7--6 and $^{13}$CO $J$=1--0, $J$=2--1, and $J$=4--3 \citep{1998A&A...331..857J,2008A&A...482..197P,2008PASA...25..129O,2008ApJS..175..485M} are also summarized in Table \ref{tbl:1}. Nine transitions are available in N159W and five in N159E and N159S.

\section{Data Analysis}\label{pasj_analysis}
\subsection{Large Velocity Gradient Analysis}\label{pasj_lvg}
In order to estimate physical properties of molecular gas, we employ a LVG model \citep{1974ApJ...189..441G}. A LVG model is useful to calculate level populations of quantum states of a molecule and line intensities of each transition. A model molecular cloud is assumed to have a large velocity gradient as compared with local velocity dispersion. The line emission from one place affects via absorption or induced emission only the local region whose size is in the order of the local velocity dispersion divided by the velocity gradient. By using an escape probability which is the probability for a once-emitted photon to escape from the cloud without re-absorption, the equations of the line radiative transfer and the statistical equilibrium are simplified and are solved locally. It is then possible to calculate the population of excitation states.

Here, we calculate level populations of $^{12}$CO and $^{13}$CO molecular rotational quantum states and line intensities under an assumption of a spherically symmetric molecular cloud of uniform density and temperature. The escape probability for a spherically symmetric velocity gradient is calculated as 
\begin{equation}
{\beta}_{J+1,J} = \frac{1-exp(-{\tau}_{J+1,J})}{{\tau}_{J+1,J}}
\end{equation}
\citep{1970MNRAS.149..111C}. The LVG model requires 3 independent physical parameters to calculate emission line intensities; kinetic temperature, gas density and $X/(dv/dr)$. $X/(dv/dr)$ is the molecular abundance divided by the velocity gradient in the cloud; [$^{12}$CO/H$_2$]/($dv/dr$) for $^{12}$CO and ([$^{12}$CO/H$_2$]/[$^{12}$CO/$^{13}$CO])/($dv/dr$) for $^{13}$CO. 

We cannot measure molecular abundances directly. Here, we estimate [$^{12}$CO/H$_2$] from the $X_{\rm CO}$-factor derived by \citet{2008ApJS..178...56F}. The $X_{\rm CO}$-factor in the LMC is about 3 times higher than the Galactic value and we shall adopt [$^{12}$CO/H$_2$] in the LMC 3 times lower than the Galactic value. Since the Galactic $X_{\rm CO}$ value was determined in the Orion molecular cloud to be $5 \times 10^{-5}$ by \citet{1987ApJ...315..621B}, the estimate [$^{12}$CO/H$_2$] in the LMC is $1.6 \times 10^{-5}$. \citet{1994A&A...291...89J} estimated that the [$^{12}$CO/$^{13}$CO] value in the N159 region is about 2/3 of the Galactic value from observations of the rare CO isotopes (C$^{18}$O, C$^{17}$O). Thus, by assuming 70 for the Galactic value we adopt [$^{12}$CO/$^{13}$CO] to be 50 in the N159 region. The velocity gradient within the molecular clumps in the N159 region is typically 0.5 km s$^{-1}$ pc$^{-1}$ \citep{2008ApJS..175..485M}. Therefore, we adopt $X/(dv/dr)=3.2 \times 10^{-5}$ for $^{12}$CO and $6.4 \times 10^{-7}$ for $^{13}$CO.

In order to obtain the optimum physical parameters, we calculate $\chi^2$ values from observed intensity ratios derived from intensities which are convolved to the same beam size (45\arcsec) at peak positions of the $^{12}$CO $J$=3--2 emission (Table \ref{tbl:2}). 

\begin{equation}
\chi^2 = \sum_{i=1}^{N-1}{\sum_{j=i+1}^{N}\frac{\{R_{obs}(i,j)-R_{LVG}(i,j)\}^2}{\sigma^2}}\\
\end{equation}
\begin{equation}
{\rm Reduced}~ \chi^2=\chi^2/(N-1)
\end{equation}
where N is the number of observed molecular transitions and ``i'' and ``j'' refer to different molecular transitions. $R_{obs}(i,j)$ is the observed line intensity ratio of transition ``i'' over transition ``j'', and $R_{LVG}(i,j)$ is the ratio of the LVG calculations at any temperature and density. ``$\sigma$'' is the uncertainty of the observed line intensity ratio. The errors in the observed intensity are mainly due to calibration uncertainty; we use conservatively $\sigma$ as 20\% of the peak line intensity ratio. Table \ref{tbl:1} shows observed line intensity at a peak position of each clump that was used in the LVG calculations. All intensities were smoothed to a 45{\arcsec} beam size.

\subsection{Results of the LVG Analysis}\label{pasj_ap_b1}
We calculated the reduced $\chi^2$ using all observed line intensities listed in Table \ref{tbl:1}. We define the range of solutions to be within the 1 $\sigma$ level of the reduced $\chi^2$ value. We display the reduced $\chi^2$ distributions in gray scale and the range of solutions in red contours in the density--temperature plane in Figure \ref{fig:4} (left). A subset of observed line intensity ratios is summarized in Table \ref{tbl:2} and also plotted in Figure \ref{fig:4} (green lines). From the minimum of reduced $\chi^2$ values in each panel, we estimate density and temperature at the three peaks as given in Table \ref{tbl:3} and find that N159W and N159E ($T_{kin} \sim $ 70 -- 80 K and $n({\rm H}_2)$ $\sim$ 3 -- 4 $\times 10^3$ cm$^{-3}$) are significantly warmer than N159S ($T_{kin} \sim $ 30 K and $n({\rm H}_2)$ $\sim$ 1.6 $\times 10^4$ cm$^{-3}$).  The reduced $\chi^2$ values of $\sim$1 for N159W and N159E indicate that the solutions are not very reliable. 

We note that locus C, the ratio between the $^{12}$CO $J$=4--3 and $J$=3--2 lines, deviates from the solution in M159W (Figures \ref{fig:4} a and b). We suggest that this can be a result of calibration error of 20\% (see Table \ref{tbl:2}).  We also note that the inclusion of loci A and B in Figures \ref{fig:4} a, c, and e, the ratios including the $^{12}$CO $J$=1--0 and $J$=2--1 transitions, tend to make the fitting somewhat worse (i.e., $\chi^2 = 1$ in N159W and N159E). We repeated the calculations by excluding intensity ratios with $^{12}$CO $J$=1--0 and $J$=2--1. The solutions are shown in Figures \ref{fig:4} b, d, and f. Thus obtained temperatures and densities are summarized in Table \ref{tbl:3}. The results are similar to those obtained for the original calculations, but the reduced $\chi^2$ value is improved (i.e. $\chi^2 = 0.5$ in N159W and $\chi^2 = 0.07$ in N159E). We suggest that the lower excitation mm emission lines may be affected by the foreground gas different from that emitting the sub-mm emission lines as shown later in Discussion.

We shall then discuss on the possible effects in determining temperature and density due to the difference in observed positions among different transitions (Table \ref{tbl:2}). The observed positions in $^{12}$CO $J$=2--1, $^{13}$CO $J$=1--0, and $^{13}$CO $J$=2--1 by \citet{1998A&A...331..857J} are shifted from those in the present work by $\sim$10 $\arcsec$ in N159W and N159E and by $\sim$30 $\arcsec$ in N159S for the 45 $\arcsec$ convolved beam. The intensity differences each between these two positions are less than 5\% according to the OTF distributions of the $^{12}$CO $J$=1--0, $J$=3--2, and $J$=4--3 lines. In LVG calculations we find the resultant differences in temperature to be $\sim$5\% in N159W and to be $\sim$10\% in N159E and N159S and that density remains the same, if the $^{12}$CO $J$=2--1, $^{13}$CO $J$=1--0 and $^{13}$CO $J$=2--1 line intensities are shifted by 10\% from the values of \citet{1998A&A...331..857J}. We summarize that the effects of the positional differences in Table \ref{tbl:2} may cause errors up to 10\% in temperature, while no significant effect in density.

\subsection{Discussions on the LVG Results} \label{chap3_dis}
We show a comparison of the calculated and observed intensities as a function of $J$ in Figure \ref{fig:5}. This comparison indicates that the line intensities in the high-excitation sub-mm regime are well reproduced and we argue that a combination of $^{12}$CO $J$=3--2, $J$=4--3, and $^{13}$CO emission lines are crucial to derive physical properties of the embedded dense region. On the other hand, in N159W and N159E the lower transitions, $^{12}$CO $J$=1--0 and $J$=2--1, are measured to be significantly below the calculated values. The $^{12}$CO $J$=1--0 line of N159W in Figure \ref{fig:3} shows a somewhat flat-top profile, suggesting possible self absorption by some foreground component which may cause the lower intensity of the $^{12}$CO $J$=1--0 emission in N159W and N159E (Figure \ref{fig:5}), while we admit the noise levels are higher in the $^{12}$CO $J$=1--0 spectra than the others. For instance, LVG calculations for density of 100 cm$^{-3}$ and temperature of 30K with the same set of the other model parameters indicate that the 20 \% decrease of the $J$=1--0 and $J$=2--1 lines in N159W is well reproduced by self-absorption by such diffuse gas, while the density and temperature of the foreground gas is not necessarily uniquely determined. Such foreground gas is not significant in emitting the $J$=3--2 and $J$=4--3 lines.

In N159S, on the other hand, all the observed intensities are consistent with calculations. 

We thus conclude that the N159W and N159E clumps consist of warm and dense gas with a temperature of 70--80 K and density of 4$\times 10^3$ cm$^{-3}$, embedded in surrounding lower density gas (n(H$_2$) $\sim $ 100 cm$^{-3}$), and that the N159S clump shows nearly uniform-temperature as is consistent with no internal significant heat source. 

A number of attempts have been made to derive physical quantities in the N159 by using multiple CO emission lines and comparing them with line transfer calculations \citep{2005ApJ...633..210B,2007A&A...471..561N,2008A&A...482..197P,2008ApJS..175..485M}. \citet{2005ApJ...633..210B} used observations of the $^{12}$CO $J$=4--3 and $J$=1--0 transitions at a 109{\arcsec} resolution and suggested that N159W has a cold component of 20 K and 10$^5$ cm$^{-3}$ and a hot component of 100 K and 10$^2$ cm$^{-3}$.
We present a comparison of the results, where we listed the most recent three works with high angular resolutions, in Table \ref{tbl:4}. The present work is distinguished from them in that it uses a combination of high excitation lines (the $J$=4--3 and $J$=3--2 lines) and optically thin $^{13}$CO lines at a high angular resolution of 45{\arcsec}, nearly equal to that of \citet{2008A&A...482..197P}. We also note that the signal to noise ratio of the present $^{12}$CO $J$=4--3 data is high (minimum S/N $\sim$ 8 at a resolution of 45{\arcsec}), helping to obtain better constraints on physical parameters.

In N159W,  \citet{2008A&A...482..197P} used $^{12}$CO $J$=4--3, $J$=7--6, $^{13}$CO $J$=4--3 and [C\emissiontype{I}] $^3P_1$--$^3P_0$, $^3P_2$--$^3P_1$ fine structure transitions observed by the NANTEN2 telescope and smoothed to a 38{\arcsec} resolution, to derive a temperature of $\sim$ 80 K and a density of $\sim$ 10$^4$ cm$^{-3}$ toward N159W, with a PDR model. \citet{2008ApJS..175..485M} used $^{12}$CO $J$=3--2, $J$=1--0 transitions and give constraints on temperature to be higher than 30 K and density in a large range from 3 $\times$ 10$^3$ cm$^{-3}$ to 8 $\times$ 10$^5$ cm$^{-3}$. The present results show a fairly good agreement with the \citet{2008A&A...482..197P}'s results. We note that the \citet{2008A&A...482..197P} observed transitions up to $^{12}$CO $J$=7--6 line, higher than the present work, which may lead to sample denser gas than the present work. We also note that the present results have significantly improved the ranges of temperature and density as compared with \citet{2008ApJS..175..485M}. A similar significant improvement is also found for N159E (Table \ref{tbl:4}). The present results for N159S are consistent with those of \citet{2008ApJS..175..485M} who used observations of $^{12}$CO and $^{13}$CO $J$=1--0 and $^{12}$CO $J$=3--2 transitions. This may reflect that N159S is a lower density cloud which can be well probed by the transitions lower than the $J$=3--2 line. A two-component model is proposed by \citet{2007A&A...471..561N} which employed the $^{12}$CO data up to the $J$=3 state and the CS data. We suggest that the model may be connected to the present idea of foreground self-absorption discussed above, while the physical parameters are yet poorly constrained.

\section{Comparisons with Star Formation Activities}\label{pasj_sf}
The N159 region is an active site of massive stare formation. The N159 itself is a group of H\emissiontype{II} regions cataloged by \citet{1956ApJS....2..315H}. \citet{1999AJ....117..238B,2008MNRAS.389..678B} and \citet{2005AJ....129..776N} presented catalogs of star clusters at optical and near infrared wavelengths. Radio studies revealed additional signs of young stars, including radio continuum thermal emission and H$_2$O and OH maser emissions \citep{1994PASAu..11...68H,1981MNRAS.194P..33C,1981Natur.290...36S}. More recently, \citet{2005ApJ...620..731J} made a study of young objects with the Spitzer IRAC, and presented a comprehensive view on young stars observable at near-- to mid--infrared wavelengths. Figures \ref{fig:6}(a) and \ref{fig:6}(b) present distributions of these young objects, stars and clusters with nebulosity. Tables \ref{tbl:5} and \ref{tbl:6} list star clusters and H\emissiontype{II} regions, respectively. The total luminosity of these objects are estimated to be $\sim 7 \times 10^7 L_{\odot}$, most of which is emitted by the most massive O type member stars. In the following, we compare the molecular gas distribution with the young star clusters and the signs of star formation activities, including mid-infrared emission. 

\subsection{H$\alpha$}
Figure \ref{fig:7} shows the distribution of the $^{12}$CO $J$=4--3 integrated intensity superposed on an H$\alpha$ image based on observations made with ESO Telescopes at the La Silla Observatory (programme ID 076.C-0888; processed and released by the ESO VOS/ADP group). The peak of N159W has no H$\alpha$ emission, but is associated with a bright, extended H$\alpha$ emission centered at (RA, Dec) $\sim$ ($5^{\rm h} 39^{\rm m} 40^{\rm s}$, -69{\arcdeg} 46{\arcmin} 30{\arcsec}) in the south. We also notice some faint H$\alpha$ features around the molecular peak of N159W. The most significant H$\alpha$ emission is extended around N159E with an extent of $\sim$ 4{\arcmin} in RA and $\sim$ 4{\arcmin} in Dec. This H\emissiontype{II} region has a dark lane extending in the east to west direction which coincides well with the N159E molecular clump. The peak molecular column densities toward N159W and N159E are estimated to be $2.2\times 10^{23}$ cm$^{-2}$ and $1.4\times 10^{23}$ cm$^{-2}$, respectively. By using the relationship with visual extinctions \citep{2007ApJ...662..969I,2008A&A...484..205D}, we obtain extinctions of 20 -- 110 mag and 18 -- 72 mag for N159W and N159E, respectively, which are large enough to explain the dark lanes by the molecular gas.

\subsection{Mid-Infrared Data from Spitzer SAGE Program}
Mid-infrared measurements are one of the useful probes for investigating star formation activities, because the spectral energy distribution (SED) peak is located in this wavelength range \citep{2007ApJ...666..870C}. Spitzer SAGE program surveyed  the entire LMC from 3.6 $\micron$ to 160 $\micron$ \citep{2006AJ....132.2268M}, allowing a study of the star formation activity throughout the LMC \citep{2008AJ....136...18W} and ``molecular ridge'' \citep{2008AJ....136.1442I}. To identify embedded young stars in the observed region, we have inspected the Spitzer data at wavelengths shorter than 70 $\micron$ \citep{2006AJ....132.2268M}, yeilding  5 mid-infrared peaks, seen as compact peaks at both 8 $\micron$ and 24 $\micron$. They are named from \#1 to \#5 as listed in Table \ref{tbl:7}. Note that this naming is different from that used by \citet{1994PASAu..11...68H}. There is an embedded ultra-compact H\emissiontype{II} region toward \#1 and another compact H\emissiontype{II} region toward \#5. Peaks \#2/\#3 and \#4 are also associated with radio H\emissiontype{II} regions. This indicates that all the five sources are associated with OB stars.

Figure \ref{fig:8} shows a comparison of the $^{12}$CO $J$=4--3 distribution with the Spitzer images at two wavelengths of 8$\micron$ and 24$\micron$. The five Spitzer sources show fairly good spatial correlation with the $^{12}$CO and are likely associated with the molecular gas. A more closer look shows that the molecular peak of N159W is significantly shifted from \#1 by $\sim$ 25{\arcsec} and from \#2 and \#3 by $\sim$ 40{\arcsec} and that the molecular peak of N159E shows an extension nearly similar to the elongated feature associated with \#5. \#4 is on the other hand located between the two $^{12}$CO clumps.

Figure \ref{fig:9} shows a comparison between the H$\alpha$ image and the Spitzer 24$\micron$ distribution. Most of the 24$\micron$ features seem well associated with the H$\alpha$. Two exceptions are in the dark lane toward N159E where H$\alpha$ becomes weak and toward the peak of N159W.  These reflect dust extinction as already noted (see Figure \ref{fig:7}). We note the two small peaks of the 24$\micron$ emission at (RA, Dec) $\sim$ ($5^{\rm h} 39^{\rm m} 30^{\rm s}$, -69{\arcdeg} 47{\arcmin} 15{\arcsec}) and ($5^{\rm h} 40^{\rm m} 00^{\rm s}$, -69{\arcdeg} 47{\arcmin} 10{\arcsec}), small star forming regions in Figure \ref{fig:6} (a), show good correspondence. 

\subsection{Comparison of the CO Line Intensity Ratio with Star Formation}
\citet{2008ApJS..175..485M} showed good correlation between line ratio $R_{3-2/1-0}$ and H$\alpha$ flux in clump-averaged values. H$\alpha$ is however not a good tracer of star formation in high density regions because of the dust extinction. Figure \ref{fig:10} shows pixel by pixel comparisons of $R_{3-2/1-0}$ which \citet{2008ApJS..175..485M} used 24$\micron$ flux. 24$\micron$ flux shows a fairly good correlation with a correlation coefficient of 0.62. Mid-infrared (MIR) emission is less sensitive to the dust extinction than shorter-wavelength emission, and is able to trace embedded young objects \citep{2005ApJ...620..731J}.  We shall compare the CO line ratio with the MIR emission obtained by Spitzer SAGE observations. 

Figure \ref{fig:11} (a) shows the distribution of the integrated intensity for the line ratio, $^{12}$CO $J$=4--3 to $^{12}$CO $J$=1--0 (hereafter $R_{4-3/1-0}$) in the northern part of N159. Figure \ref{fig:11} (b) shows the same distribution, superposed  on contours of the 24$\micron$ emission observed with Spitzer. In general, the $^{12}$CO $J$=4--3 emission shows a good correlation with $R_{4-3/1-0}$. Figure \ref{fig:11} (b) reveals that MIR peak \#5 and its associated extension at 24$\micron$ in N159E are in excellent agreement with the enhancement of $R_{4-3/1-0}$, suggesting that the heating by the 24$\micron$ source leads to the enhanced $R_{4-3/1-0}$. 24$\micron$ peaks in N159W, \#1 and \#2/\#3, are located toward the edge of the enhancement of $R_{4-3/1-0}$ and a similar trend is found with $R_{3-2/1-0}$ (Figure \ref{fig:11}(c)). The region of high line ratios with $R_{4-3/1-0}$ greater than 1.0 is distributed in the $^{12}$CO $J$=4--3 clump of N159W and shows no sign of star formation except for the three MIR sources. By considering that the high line ratios indicate higher density, this motivates us to suggest that the main part of the N159W cloud is prior to active star formation except for the three MIR sources located toward the eastern edge and a few faint H\emissiontype{II} regions. Further detailed observations at higher resolutions of the N159W cloud should be of considerable importance in probing the initial conditions of massive star formation.

Figure \ref{fig:11} (d) shows that OB stars and Herbig Ae/Be stars are mainly distributed between the two molecular peaks, where the $J$=4--3 emission is weak. This suggests that the natal molecular gas of the young stars has been dissipated by the ionization and the dynamical effects of the stellar winds. 

We show the same comparison for N159S as shown in Figure \ref{fig:12}. We find a slight enhancement of $R_{4-3/1-0}$ but no significant hint of star formation. The marginal enhancement in $R_{4-3/1-0}$ in the east is ascribed to a small H\emissiontype{II} region \citep{2008ApJS..175..485M}.

\subsection{Star Formation in N159 and the Nature of the N159W Molecular Core}
N159 is one of the most active sites of star formation in the LMC \citep{Kawamura09}. The present study combined with the $^{12}$CO $J$=3--2 study by \citet{2008ApJS..175..485M} and \citet{2008A&A...482..197P} provides a better insight into temperature and density in the N159 molecular clumps by using the sub--mm transitions of CO at 10 pc resolution.

The works on star clusters cited in Table \ref{tbl:5} indicate that the N159 region is similar to the Galactic massive star forming regions, in terms of the level of activity and short time scale \citep{2005ApJ...620..731J,1999AJ....117..238B,2005AJ....129..776N}. The main properties of star formation are summarized in Table \ref{tbl:8}, along with those with the $\eta$ Car region for comparison. 

We make a comparison of the N159 GMC and $\eta$ Car GMC, one of the most active massive star forming regions in the Galaxy. In Figure \ref{fig:13} the top-left diagram shows molecular distribution of N159W observed in the $^{12}$CO $J$=3--2 emission line and the bottom-left shows that observed in the $^{12}$CO $J$=1--0 emission line. The two distributions are converted into the same spatial resolution with the top diagram in the $\eta$ Car northern cloud\citep{2005ApJ...634..476Y}. While $^{12}$CO $J$=1--0 and $J$=3--2 emission lines trace different excitation conditions, the distributions of these lines are similar with each other qualitatively as mentioned in Section \ref{chap3_res1}. We also show the intensity distributions along with the X and Y axes in top-right and bottom-right diagrams.

Observational properties of star formation activity such as the number of O stars and IR luminosities and the properties of GMCs, such as sizes and masses, in the N159 region and $\eta$ Car region are summarized in Table \ref{tbl:8}. The total mass and size of the two cores are nearly the same, $\sim 10^5 M_{\odot}$ and $\sim 10$ pc. The star clusters are mostly OB associations including 4--10 O3--O5 stars over an area of $\sim$ 100 pc$^2$ and there is no sign of spatially confined super star cluster like R136. 

N159 is Type III in the GMC classification \citep{Kawamura09}. The present work at higher resolution indicates that a GMC may be resolved into sub structures; clumps N159W, N159E and N159S. Such clumps may be classified in a similar way from Type I to III; N159W and N159E are Type III, and N159S Type I or II as has already been discussed by \citet{2008ApJS..175..485M}. Future follow-up works at high resolutions will help to refine this classification.

The present study has shown that the western part of the N159W core is not associated with any optical or infrared features, suggesting that it is at a stage prior to active star formation. Further detailed observations at higher resolutions should be of considerable importance in probing the initial conditions of massive star formation.

\section{Summary}\label{pasj_sum}
We carried out $^{12}$CO $J$=4--3 observations of the N159 region with NANTEN2 and $^{13}$CO $J$=3--2 observations toward $^{12}$CO $J$=3--2 peaks with ASTE. The main results are summarized as follows;

\begin{enumerate}
\item The N159 GMC has been resolved into three prominent clumps, N159W, N159E, and N159S in the $^{12}$CO $J$=4--3 emission line. N159W shows the highest intensities among the three in the CO transitions from $J$=1--0 to $J$=7--6.

\item Molecular densities and temperatures have been derived toward the three peaks. Using a LVG analysis involving the $^{12}$CO $J$=4--3 , $^{13}$CO $J$=3--2 emission lines in addition to the other CO lines published, we find that N159W and N159E have temperature of $\sim$ 70 -- 80 K and density of $\sim 3 \times 10^3$ cm$^{-3}$, and that N159S has temperature of $\sim$ 30 K and density of $\sim 1.6 \times 10^3$ cm$^{-3}$. In order to explain lower line intensities than expected, we suggest that $^{12}$CO $J$=1--0 and $J$=2--1 lines may be affected by self-absorption by foreground lower excitation gas.

\item The $^{12}$CO $J$=4--3 distribution is compared with H$\alpha$ and mid-- to far--infrared emission, a sign of embedded star formation, obtained with the Spitzer SAGE program. N159E shows a good coincidence with a dark lane of H$\alpha$ and also with a 24 $\micron$ extended source. On the other hand, N159W is associated with three compact 24 $\micron$ sources and some small H$\alpha$ features, although the $^{12}$CO $J$=4--3 peak of N159W and its western part show no sign of star formation.

\item A comparison between N159 and $\eta$ Car indicates that they show similar star formation activity and we do not see significant difference in physical parameters between these two massive star forming regions.  
\end{enumerate} 

 \bigskip
We thank the all members of the NANTEN2 consortium and ASTE team for the operation and persistent efforts to improve the telescopes. 
This research was supported by the Grant-in-Aid for Nagoya
University Global COE Program, "Quest for Fundamental Principles in the
Universe: from Particles to the Solar System and the Cosmos", from the Ministry
of Education, Culture, Sports, Science and Technology of Japan.
This work is financially supported in part by a Grant-in-Aid for Scientific Research from the Ministry of Education, Culture, Sports, Science and Technology of Japan (No. 15071203) and from JSPS (Nos. 14102003 and 18684003), and by the JSPS core-to-core program (No. 17004), and the Mitsubishi Foundation. 
This work is also financially supported in part by the grant SFB 494 of the Deutsche Forschungsgemeinschaft, the Ministerium fur Innovation, Wissenschaft, Forschung und Technologie des Landes Nordrhein-Westfalen and through special grants of the Universitat zu Koln and Universitat Bonn. 
SAGE research has been funded by NASA/Spitzer grant 1275598 and NASA NAG5-12595. 
The ASTE project is driven by Nobeyama Radio Observatory (NRO), a branch of National Astronomical Observatory of Japan (NAOJ), in collaboration with University of Chile, and Japanese institutes including University of Tokyo, Nagoya University, Osaka Prefecture University, Ibaraki University., and Hokkaido University. Observations with ASTE were in part carried out remotely from Japan by using NTT's GEMnet2 and its partner R\&E (Research and Education) networks, which are based on AccessNova collaboration of University of Chile, NTT Laboratories, and NAOJ. A part of this study was financially supported by the MEXT Grant-in-Aid for Scientific Research on Priority Areas No.\ 15071202.

\clearpage

\begin{table}
  \begin{center}
     \caption{Observed intensity peak properties} 
     \label{tbl:1}
     \begin{tabular}{rlcccccc}
      \hline
 \multicolumn{2}{c}{Line}          &\multicolumn{2}{c}{Position(J2000)}& {$T_{mb}$} & {$V_{LSR}$} &{$dV$} &{ref} \\
        &     & R.A. (h m s)& Dec. ({\arcdeg} {\arcmin} {\arcsec}) & (K) & (\mbox{km s$^{-1}$}) & (\mbox{km s$^{-1}$}) \\
      \hline
\multicolumn{1}{l}{N159W}\\
$^{12}$CO&$J$=1--0 & 5 39 37.3& -69 45 34.5 & 5.8 & 237.7 & 7.4  & 1\\
         &$J$=2--1 & 5 39 35.1& -69 45 24.8 & 6.3  & 237.6 & 8.2  & 2\\
         &$J$=3--2 & 5 39 36.8& -69 45 31.9 &14   & 237.3 & 8.1  & 3\\
         &$J$=4--3 & 5 39 36.3& -69 45 44.5 & 8.2 & 237.5 & 7.1  & 4\\
         &$J$=7--6 & 5 39 36.8& -69 45 31.9 & 3.3  & -     & -    & 5 \\
$^{13}$CO&$J$=1--0 & 5 39 35.1& -69 45 24.8 & 0.80 & 237.6 & 7.8  & 2 \\
         &$J$=2--1 & 5 39 35.1& -69 45 24.8 & 1.4  & 237.4 & 7.4  & 2 \\
         &$J$=3--2 & 5 39 36.8& -69 45 31.9 & 3.1 & 238.0 & 7.4  & 4\\
         &$J$=4--3 & 5 39 36.8& -69 45 31.9 & 0.95 & -     & -    & 5\\
\hline
\multicolumn{1}{l}{N159E}\\
$^{12}$CO&$J$=1--0 & 5 40 9.3& -69 44 24.5 & 3.8 & 234.2 & 6.5  & 1 \\
         &$J$=3--2 & 5 40 8.7& -69 44 34.3 & 8.9  & 233.2 & 7.0  & 3\\
         &$J$=4--3 & 5 40 7.3& -69 44 54.6 & 5.1 & 233.6 & 7.0  & 4\\
$^{13}$CO&$J$=1--0 & 5 40 8.4& -69 44 47.2 & 0.44 & 234.1 & 6.6  & 2 \\
         &$J$=3--2 & 5 40 8.7& -69 44 34.3 & 1.5 & 233.6 & 5.4 &  4\\
\hline
\multicolumn{1}{l}{N159S}\\
$^{12}$CO&$J$=1--0 & 5 40 4.6& -69 50 34.5 & 4.6 & 235.1 & 8.3 &  1 \\
         &$J$=3--2 & 5 40 5.0& -69 50 34.0 & 4.1  & 234.4 & 9.7 & 3\\
         &$J$=4--3 & 5 40 6.5& -69 50 44.5 & 2.3 & 234.3 & 9.0 &  4\\
$^{13}$CO&$J$=1--0 & 5 39 58.8& -69 50 29.6 & 0.72 & 235.5 & 5.9 & 2\\
         &$J$=3--2 & 5 40 5.0& -69 50 34.0 & 0.52 & 237.0 & 4.6 &  4\\

\hline
\ \\
\multicolumn{7}{@{}l@{}}{\hbox to 0pt{\parbox{133mm}{
  \par\noindent
  Reference --- (1) \cite{2008PASA...25..129O}; (2) \cite{1998A&A...331..857J}; (3) \cite{2008ApJS..175..485M} Beam efficiency of this observation was reviced down to 0.6 from 0.7; (4) This work; (5) \cite{2008A&A...482..197P}.
  \par\noindent
}\hss}}

    \end{tabular}
  \end{center}
\end{table}

\begin{table}[t]
  \begin{center}
     \caption{Intensity and intensity ratio at clump peak} 
     \label{tbl:2}
     \begin{tabular}{rlcccc}
      \hline
 & & N159W & N159E & N159S & \\
     \hline
\multicolumn{2}{l}{Peak position}\\
\multicolumn{2}{c}{R.A. (J2000) ($^{\rm h~ m~ s}$)} & 5 39 36.8 & 5 40 8.7 & 5 40 5.0\\
\multicolumn{2}{c}{DEC. (J2000) ({\arcdeg} {\arcmin} {\arcsec})} & -69 45 31.9 & -69 44 34.3 & -69 50 34.0 \\
     \hline
\multicolumn{2}{l}{45{\arcsec} beam intensity (K)}\\     
 $^{12}$CO&$J$=1--0 & 5.9& 3.9& 4.6& \\
         &$J$=2--1\footnotemark[$*$] & 6.3& -& -& \\
         &$J$=3--2 & 7.7& 5.4& 3.7& \\
         &$J$=4--3 & 7.5& 4.5& 2.2& \\
         &$J$=7--6 & 2.5& -& -& \\
 $^{13}$CO&$J$=1--0\footnotemark[$*$] & 0.80& 0.44& 0.72& \\
         &$J$=2--1\footnotemark[$*$] & 1.4& -& -& \\
         &$J$=3--2 & 1.3& 0.86& 0.27& \\
         &$J$=4--3 & 0.89& -& -& \\
      \hline
\multicolumn{2}{l}{Line intensity ratio}  &   &   &   & Mark of figure\\

\multicolumn{2}{c}{$^{12}$CO $J$=4--3/$^{12}$CO $J$=1--0} & 1.3 $\pm$ 0.3  & 1.2 $\pm$ 0.2  & 0.48 $\pm$ 0.1   & A\\
\multicolumn{2}{c}{$^{12}$CO $J$=4--3/$^{12}$CO $J$=2--1} & 1.2 $\pm$ 0.2  & -              & -  & B\\
\multicolumn{2}{c}{$^{12}$CO $J$=4--3/$^{12}$CO $J$=3--2} & 0.97$\pm$ 0.19 & 0.83 $\pm$ 0.17& 0.59 $\pm$ 0.1  & C\\
\multicolumn{2}{c}{$^{12}$CO $J$=7--6/$^{12}$CO $J$=4--3} & 0.33$\pm$ 0.06 & -                     & -                      & D\\
\multicolumn{2}{c}{$^{12}$CO $J$=4--3/$^{13}$CO $J$=1--0} & 9.4 $\pm$ 1.9  & 10 $\pm$ 2 & 3.1 $\pm$ 0.6    & E\\
\multicolumn{2}{c}{$^{12}$CO $J$=4--3/$^{13}$CO $J$=2--1} & 5.4 $\pm$ 1.1  & -                     & -                      &F\\
\multicolumn{2}{c}{$^{12}$CO $J$=4--3/$^{13}$CO $J$=3--2} & 5.8 $\pm$ 1.2  & 5.2 $\pm$ 1.0 & 8.1 $\pm$ 1.6   & G \\
\multicolumn{2}{c}{$^{12}$CO $J$=4--3/$^{13}$CO $J$=4--3} & 8.4 $\pm$ 1.7  & -                     & -                      & H\\
\hline
\ \\
\multicolumn{6}{@{}l@{}}{\hbox to 0pt{\parbox{140mm}{
  \par\noindent
  Note.-- Convolved intensity and their ratio of 45{\arcsec} resolution at $^{12}$CO $J$=3--2 integrated intensity peak position. Intensities of $^{13}$CO $J$=1--0 and $^{13}$CO $J$=2--1 are at $^{12}$CO $J$=1--0 integrated intensity peak by \cite{1998A&A...331..857J}.\\
  \footnotemark[$*$] Observed position is different. N159W of $5^{\rm h} 39^{\rm m} 35.1^{\rm s}$, -69{\arcdeg} 45{\arcmin} 24.8{\arcsec}, N159E of $5^{\rm h} 40^{\rm m} 8.4^{\rm s}$, -69{\arcdeg} 44{\arcmin} 47.2{\arcsec}, N159S of $5^{\rm h} 39^{\rm m} 58.8^{\rm s}$, -69{\arcdeg} 50{\arcmin} 29.6{\arcsec}. 
}\hss}}
    \end{tabular}
  \end{center}
\end{table}

\begin{table}[t]
  \begin{center}
     \caption{The results of the LVG calculation in each clump} 
     \label{tbl:3}
     \begin{tabular}{cccc}
      \hline
        & ~~N159W~~ & ~~N159E~~ & ~~N159S~~\\
      \hline
\multicolumn{1}{l}{One component}\\
  Reduced $ \chi ^2$ & 1.05 & 0.95 & 0.11\\
  $T_{kin} $ (K)                      & 69 $ ^{+3} _{-3} $  & 79 $ ^{+12} _{-10} $     & 31 $ ^{+8} _{-9} $ \\
  $n(H_2) (10^3$ cm$^{-3})$ & 4.0 $ ^{+0.0} _{-0.0} $ & 4.0 $ ^{+1.0} _{-0.0} $ & 1.6 $ ^{+0.4} _{-0.3} $ \\
  $N_{\rm CO} (10^{18}$ cm$^{-2})$ & 3.5 $ ^{+0} _{-0} $ & 2.9 $ ^{+0.7} _{-0} $ & 1.2 $ ^{+0.3} _{-0.3} $ \\
\hline
\multicolumn{1}{l}{High excitation component}\\
  Reduced $ \chi ^2$ & 0.52 & 0.07 & 0.10\\
  $T_{kin} $ (K)                      & 72 $ ^{+9} _{-9} $  & 83$ ^{+26} _{-20} $    & 31 $ ^{+10} _{-8} $ \\
  $n(H_2) (10^3$cm$^{-3})$ & 4.0 $ ^{+0} _{-0.8} $ & 3.1 $ ^{+1.9} _{-0.6} $ & 1.6 $ ^{+0.4} _{-0.3} $ \\
  $N_{\rm CO} (10^{18}$ cm$^{-2})$ & 3.5 $ ^{+0} _{-0.7} $ & 2.3 $ ^{+1.4} _{-0.5} $ & 1.3 $ ^{+0.3} _{-0.3} $ \\
\hline
\ \\
\multicolumn{4}{@{}l@{}}{\hbox to 0pt{\parbox{100mm}{
  \par\noindent
  Note.-- Calculated temperature and density from LVG simulation
  where $\chi^2$ value is minimum. $N_{\rm CO}$ is calculated using averaged
  velocity dispersion derived by $^{12}$CO $J$=3--2 observation of
  \cite{2008ApJS..175..485M}.
}\hss}}
   \end{tabular}
  \end{center}
\end{table}

\begin{table}[t]
  \begin{center}
     \caption{Summary of excitation analysis} 
     \label{tbl:4}
     \begin{tabular}{ccccccccc}
      \hline
 & & & \multicolumn{2}{c}{N159W} & \multicolumn{2}{c}{N159E} & \multicolumn{2}{c}{N159S} \\
\multicolumn{2}{c}{Paper} & ref. &$T_{kin}$ & $n(H_2)$ & $T_{kin}$ & $n(H_2)$ & $T_{kin}$ & $n(H_2)$ \\
 & & &(K) & (10$^3$ cm$^{-3}$) & (K) & (10$^3$ cm$^{-3}$) & (K) & (10$^3$ cm$^{-3}$) \\
\hline\\
\multicolumn{2}{c}{This work} & & 72 $^{+9}_{-9}$ & 4.0 $^{+0}_{-0.8}$ & 83$ ^{+26} _{-20} $  & 3.1 $ ^{+1.9} _{-0.6} $  & 31 $ ^{+10} _{-8} $ & 1.6 $ ^{+0.4} _{-0.3} $\\
Nikolic07 &CDC &(1)& 10 - 100 & $>$ 1 & --- & --- & $<20$ & all\\
          &HTC &(2)& $>$100 & $>$3 & --- & --- & $>20$ & $<3$\\ 
\multicolumn{2}{c}{Minamidani08} &(3)&$>$30 & 3 - 800 & $>$40 & 1 - 300 & 20 - 60 & 1 - 6\\
\multicolumn{2}{c}{Pineda08} &(4)&80 & 10 - 100 & --- & --- & --- & ---\\

\hline
\ \\
\multicolumn{8}{@{}l@{}}{\hbox to 0pt{\parbox{145mm}{
  \par\noindent
  Reference --- (1) Hot component of results from RADEX calculation by \citet{2007A&A...471..561N}. (2) Cold component of calculation by \citet{2007A&A...471..561N}. (3) LVG analysis by \citet{2008ApJS..175..485M}. (4) Clumpy-PDR analysis by \citet{2008A&A...482..197P}.
}\hss}}
    \end{tabular}
  \end{center}
\end{table}

\begin{table}[t]
  \begin{center}
     \caption{Cluster catalog} 
     \label{tbl:5}
     \begin{tabular}{ccccccc}
      \hline
     & \multicolumn{2}{c}{Position(J2000)} &  \multicolumn{2}{c}{Size}  & & Associated \\    
 object & R.A. & DEC.  & maj & min & ref & Molecular clump\\
 & (h m s) & ({\arcdeg}~{\arcmin}~{\arcsec}) &  (\arcmin)&  (\arcmin)\\
   \hline
LMC-N159H  & 5 39 30   & -69 47 28   & 0.3 & 0.3 & 1 & N159W\\
LMC-N159K  & 5 39 31   & -69 46  5   & 0.3 & 0.3 & 1 & N159W\\
LMC-N159J  & 5 39 32   & -69 43 54   & 0.3 & 0.3 & 1 & N159W\\
LMC-N159A1 & 5 39 39   & -69 45 52   & 0.3 & 0.3 & 1 & N159W\\
NGC2079    & 5 39 39   & -69 46 22   & 1.1 & 0.9 & 1 & N159W\\
NGC2078    & 5 39 40   & -69 44 21   & 1.2 & 1.1 & 1 & N159W\\
LMC-N159F  & 5 39 40   & -69 44 33   & 0.8 & 0.7 & 1 & N159W\\
LMC-N159A2 & 5 39 40   & -69 46 56   & 0.3 & 0.3 & 1 & N159W\\
NGC2084w   & 5 39 53   & -69 45 42   & 1.5 & 1.3 & 1 & N159E\\
BSDL2727   & 5 39 54   & -69 43 22   & 0.4 & 0.4 & 1 & -\\
NGC2083    & 5 39 59   & -69 44  9   & 2.0 & 1.8 & 1 & N159E\\
LMC-N159E  & 5 40  2   & -69 47 12   & 0.6 & 0.6 & 1 & -\\
NGC2084e   & 5 40  6   & -69 45 51   & 1.7 & 1.2 & 1 & N159E\\
BSDL2755   & 5 40 19   & -69 46 26   & 0.8 & 0.6 & 1 & -\\
N159-Y4    & 5 40  4   & -69 45  8   & 1.1 & 1.1 & 2 & N159E\\
LMC-N159L  & 5 40  2   & -69 49 10   & 1.4 & 1.1 & 1 & N159S\\
LMC-N173   & 5 40 20   & -69 53  8   & 0.9 & 0.9 & 1 & N159S\\
LMC-DEM279 & 5 40 26   & -69 50 22   & 1.2 & 1.2 & 1 & -\\
N159S-Y1   & 5 40 13   & -69 49 57   & 0.6 & 0.6 & 2 & N159S\\

\hline
\ \\
\multicolumn{6}{@{}l@{}}{\hbox to 0pt{\parbox{120mm}{
  \par\noindent
  Reference --- (1) \cite{1999AJ....117..238B}; (2) \cite{2005AJ....129..776N}.
}\hss}}
    \end{tabular}
  \end{center}
\end{table}

\begin{table}[t]
  \begin{center}
     \caption{HII regions} 
     \label{tbl:6}
     \begin{tabular}{ccccccccc}
      \hline
           & \multicolumn{2}{c}{Position(J2000)}&\multicolumn{2}{c}{Size} & & Ionizing & & Associated \\    
 object  & R.A. & DEC. & maj & min &ref & Source & ref &Molecular clump\\
                   & (h m s) & ({\arcdeg}~{\arcmin}~{\arcsec}) & (\arcsec) & (\arcsec)\\
   \hline
LMC-N159H  & 5 39 30    & -69 47 28    &  -  &  -  & 1 & - & - & N159W \\
LMC-N159K  & 5 39 31    & -69 46  5    &  25 &  22 & 1 & - & - & N159W \\
LMC-N159J  & 5 39 32    & -69 43 54    &  -  &  -  & 1 & - & - & N159W \\
LMC-N159A1 & 5 39 39    & -69 45 52    &  -  &  -  & 2 & - & - & N159W \\
LMC-N159A  & 5 39 39    & -69 46 22    &  56 &  62 & 1 & O5V+O7V & 3 & N159W\\ 
LMC-N159F  & 5 39 40    & -69 44 33    &  48 &  47 & 1 & - & - & N159W \\ 
LMC-N159A2 & 5 39 40    & -69 46 56    &  -  &  -  & 2 & O9V & 3 & N159W \\
LMC-N159C  & 5 39 53    & -69 45 42    & 140 & 107 & 1 & - & - & N159E \\ 
LMC-N159D  & 5 39 59    & -69 44  9    & 131 &  97 & 1 & - & - & N159E \\ 
LMC-N159E  & 5 40  2    & -69 47 12    &  35 &  35 & 1 & - & - & -     \\
LMC-N159L  & 5 40  2    & -69 49 10    &  62 &  28 & 1 & - & - & N159S \\
LMC-N173   & 5 40 20    & -69 53  8    &  49 &  54 & 1 & - & - & N159S \\
LMC-DEM279 & 5 40 26    & -69 50 22    &  65 &  65 & 4 & - & - & -     \\
N159AN     & 5 39 37    & -69 45 26    & 3.6 & 3.1 & 5 & 2$\times$O4V/O5V & 6,7 & N159W\\
Papillon   & 5 40  5    & -69 44 38    & 3.1 & 2.7 & 8 & O4V/O5V & 6 & N159E \\ 
\hline
\ \\
\multicolumn{9}{@{}l@{}}{\hbox to 0pt{\parbox{120mm}{
  \par\noindent
  Reference --- (1) \cite{1956ApJS....2..315H}; (2) \cite{1982A&A...111L..11H}; (3) \cite{1992A&A...259..480D}; (4) \cite{1976MmRAS..81...89D}; (5) \cite{1994PASAu..11...68H}; (6) \cite{2005A&A...433..205M}; (7) \cite{2003ApJ...596L..83I}; (8) \cite{1999A&A...352..665H}.
}\hss}}
    \end{tabular}
  \end{center}
\end{table}

\begin{table}[t]
  \begin{center}
     \caption{Mid-infrared peaks} 
     \label{tbl:7}
     \begin{tabular}{cccccc}
      \hline
        &    \multicolumn{2}{c}{Peak position (J2000)} & Associated & Associated & IR \\
 Peak num.& R.A.& DEC. & Molecular clump & H II region & Luminosity \\
        & (h m s) & ({\arcdeg}~{\arcmin}~{\arcsec}) &  & &($\times 10^5 L_{\odot}$)\\
      \hline
 \#1 & 5 39 38 & -69 45 26 &N159W& N159AN, HW\footnotemark[$\dagger$]\#4 & 3.7\\
 \#2 & 5 39 37 & -69 46  8  &N159W& N159A, HW\#5  & 1.7\\
 \#3 & 5 39 42 & -69 46 11 &N159W& N159A, HW\#5          & 1.3\\
 \#4 & 5 39 52 & -69 45 17 &-& HW\#2/\#3                  & 1.2\\
 \#5 & 5 40 05 & -69 44 38 &N159E& Papillon nebula, HW\#1 & 3.4\\
 \hline
\ \\
\multicolumn{6}{@{}l@{}}{\hbox to 0pt{\parbox{120mm}{
  \par\noindent
  \footnotemark[$\dagger$] HW means \cite{1994PASAu..11...68H}.
}\hss}}
    \end{tabular}
  \end{center}
\end{table}

\begin{table}[t]
  \begin{center}
     \caption{Comparisons with $\eta$ Carinae GMC} 
     \label{tbl:8}
     \begin{tabular}{ccccc}
      \hline
 & N159 region & ref & $\eta$ Car region & ref \\
      \hline
Massive star &  3$\times$ O4/O5 & 1 & 5$\times$O3+$\eta$Car& 2\\

Num. of O star & $\geq $ 55 & 3 & 70 & 4 \\

Size of GMC \\
(pc$\times$pc)& 130$\times$90&5 &150$\times$100& 6\\

Mass of GMC \\
($\times 10^5 M_{\odot}$) &5.3 &5 &3.4&6\\

Total IR luminosity\\
($\times 10^7 L_{\odot}$) & $\sim 0.9$ &7 &  1.2       & 8\\
\hline
\ \\
\multicolumn{5}{@{}l@{}}{\hbox to 0pt{\parbox{105mm}{
  \par\noindent
  Reference --- (1) Estimated from $N_{LYC}$ derived by \cite{2003ApJ...596L..83I} and \cite{2005A&A...433..205M}; (2) \cite{2006MNRAS.367..763S}; (3) \citet{2005AJ....129..776N}; (4) Simbad astronomical database; (5) $M_{\rm vir}$ from $^{12}$CO $J$=3--2 observation by \cite{2008ApJS..175..485M}; (6) $M_{\rm CO}$ from $^{12}$CO $J$=1--0 observation by \cite{2005ApJ...634..476Y};(7) \cite{2005ApJ...620..731J};  (8) \cite{2007MNRAS.379.1279S}.
}\hss}}
    \end{tabular}
  \end{center}
\end{table}

\clearpage

\clearpage

\begin{figure}
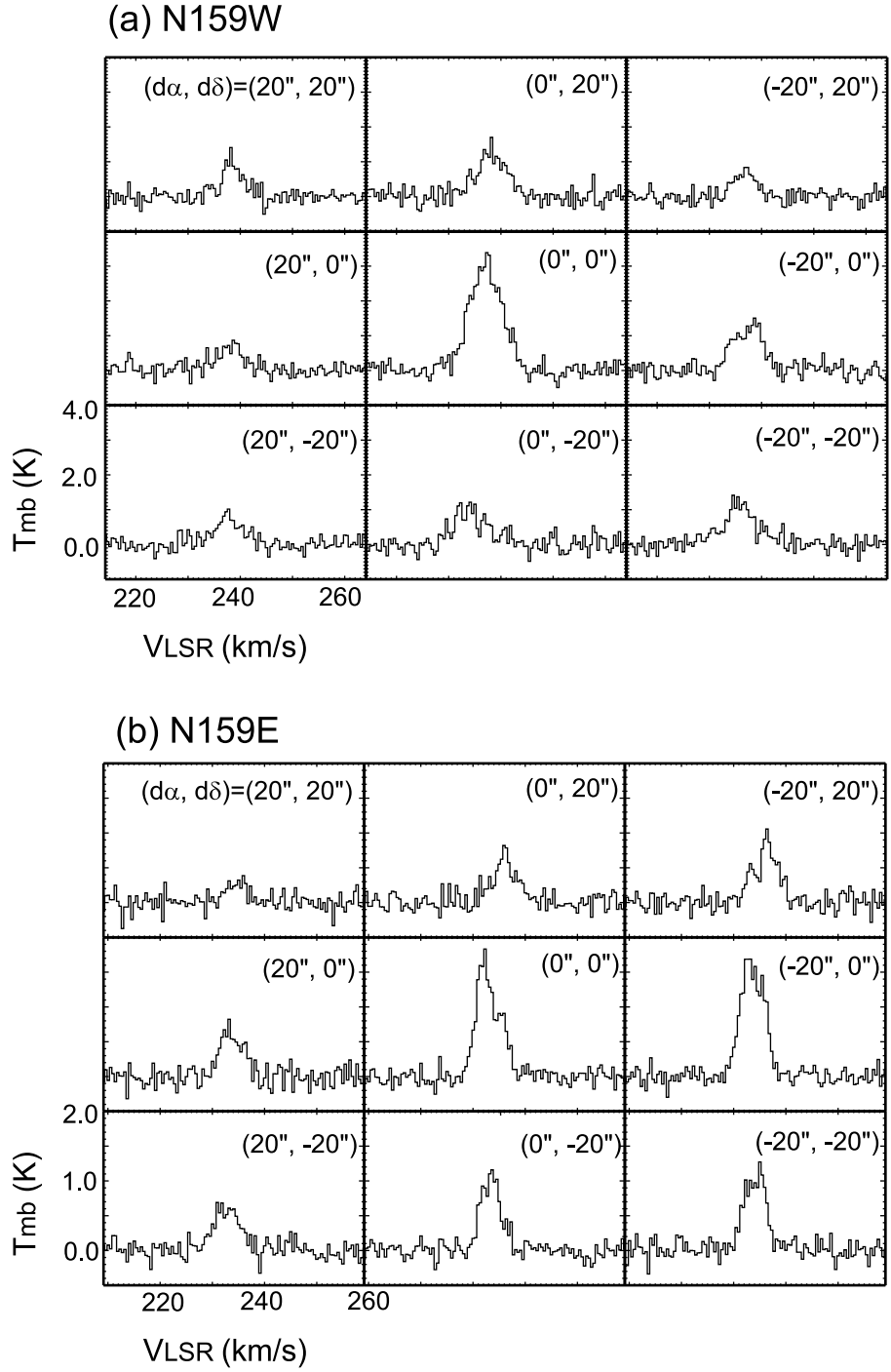

\begin{center}
\FigureFile(120mm,80mm){figure2a.eps}
\caption{Profile map of $^{13}$CO $J$=3--2. (a) N159W (center position is $5^{\rm h} 39^{\rm m} 36.8^{\rm s}$, -69{\arcdeg} 45{\arcmin} 32{\arcsec} at J2000 coordinate), (b) N159E ($5^{\rm h} 40^{\rm m} 8.7^{\rm s}$, -69{\arcdeg} 44{\arcmin} 34{\arcsec}), and (c) N159S ($5^{\rm h} 40^{\rm m} 5^{\rm s}$, -69{\arcdeg} 50{\arcmin} 34{\arcsec}).}
\label{fig:2}
\end{center}
\end{figure}
\clearpage

\setcounter{figure}{1}
\begin{figure}
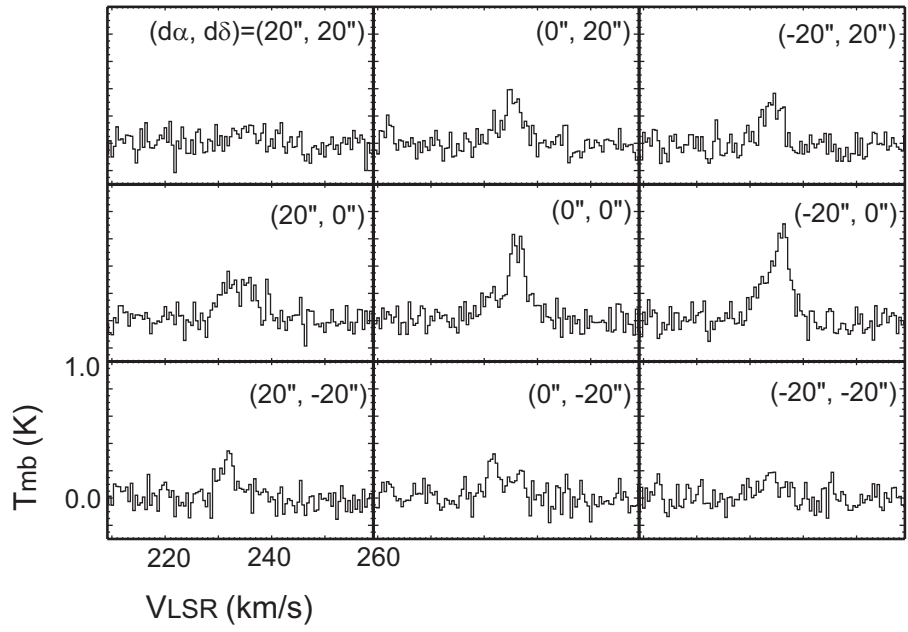

\begin{center}
\FigureFile(120mm,80mm){figure2b.eps}
\caption{{\it continued}}
\end{center}
\end{figure}
\clearpage

\begin{figure}
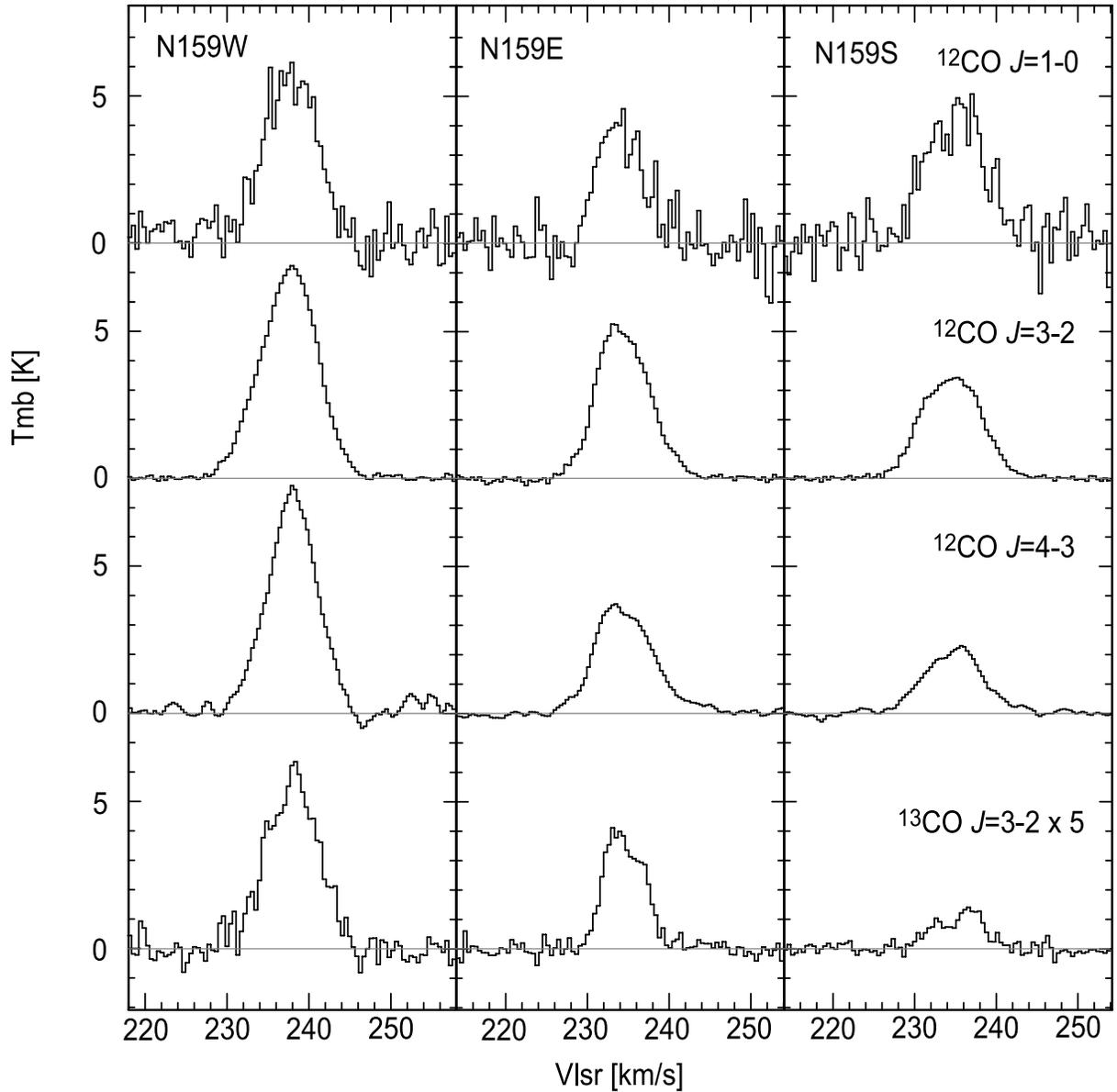

\begin{center}
\FigureFile(160mm,160mm){figure3.eps}
\caption{$^{12}$CO $J$=1--0, $^{12}$CO $J$=3--2, $^{12}$CO $J$=4--3 and $^{13}$CO $J$=3--2 line spectra from peak position of N159W (left), N159E (middle) and N159S (right). All spectra were convolved to 45{\arcsec} resolution. $^{13}$CO $J$=3--2 spectra were multiplied by 5.
}
\label{fig:3}
\end{center}
\end{figure}
\clearpage

\begin{figure}
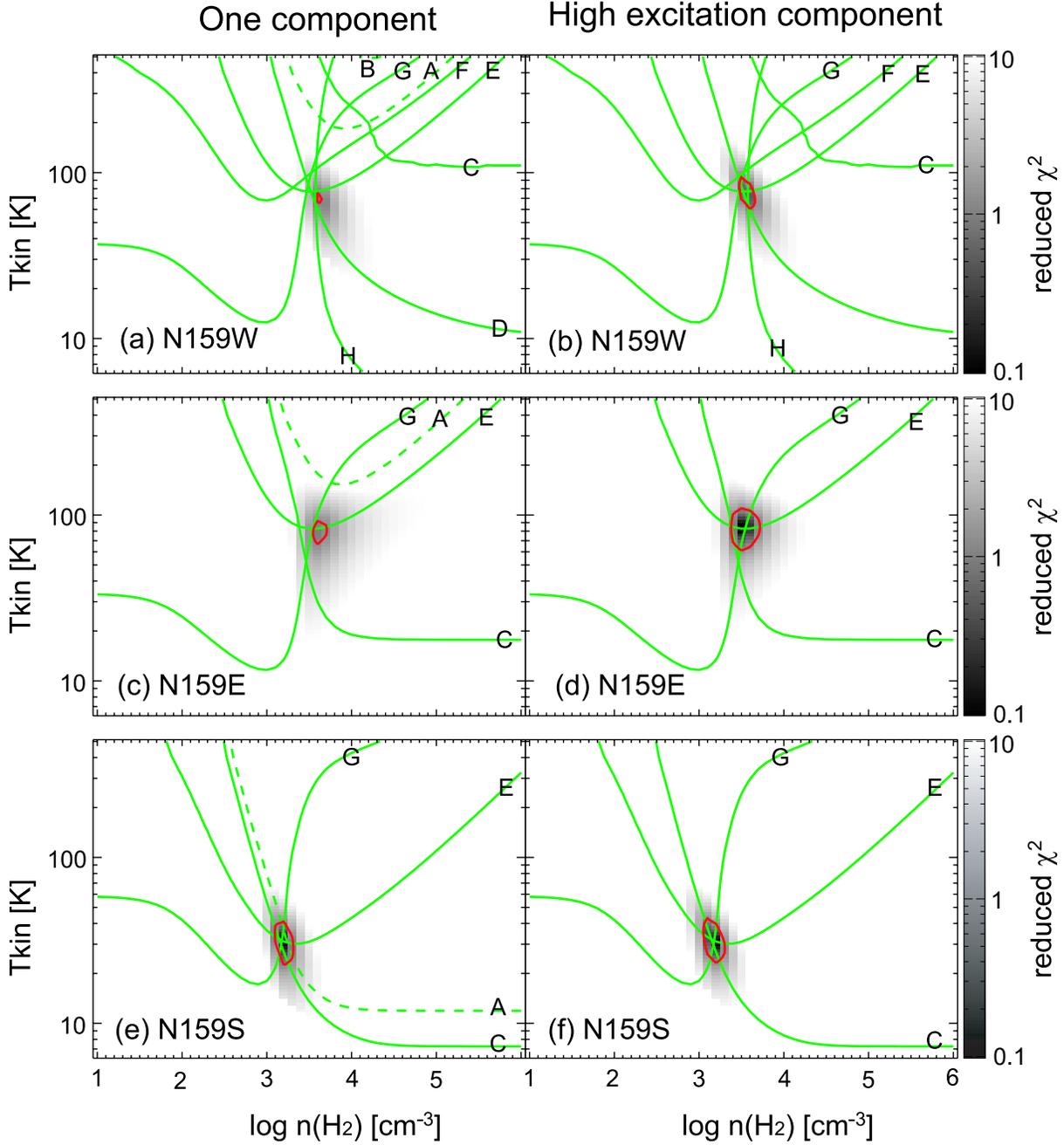

\begin{center}
\FigureFile(160mm,160mm){figure4.eps}
\caption{Observed line intensity ratio overlaid with distributions of the reduced $\chi^2$ value. Green lines shows a subset of line ratios.  Alphabets correspond column (5) of Table \ref{tbl:2}; (A) $^{12}$CO $J$=4--3/$^{12}$CO $J$=1--0, (B) $^{12}$CO $J$=4--3/$^{12}$CO $J$=2--1, (C) $^{12}$CO $J$=4--3/$^{12}$CO $J$=3--2, (D) $^{12}$CO $J$=7--6/$^{12}$CO $J$=4--3, (E) $^{12}$CO $J$=4--3/$^{13}$CO $J$=1--0, (F) $^{12}$CO $J$=4--3/$^{13}$CO $J$=2--1, (G) $^{12}$CO $J$=4--3/$^{13}$CO $J$=3--2, (H) $^{12}$CO $J$=4--3/$^{13}$CO $J$=4--3. Red contour means threshold value of $\chi^2$. Left diagrams are the results of LVG simulation using all observed lines. Right diagrams are the result using $^{12}$CO $J$=3--2 and higher excited rotational transitions and all $^{13}$CO lines.
}
\label{fig:4}
\end{center}
\end{figure}
\clearpage

\begin{figure}
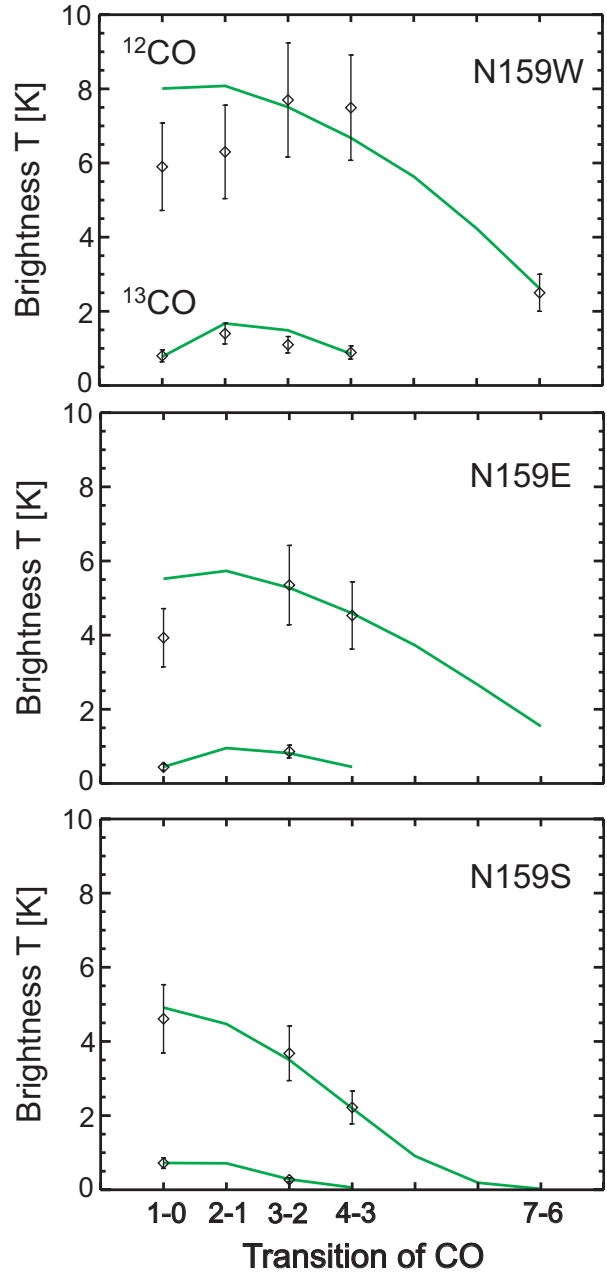

\begin{center}
\FigureFile(80mm,160mm){figure5.eps}
\caption{Plots show calculated intensity from LVG with best fit parameters of high excitation component (green line) and observed line intensities (mark with error bar).}
\label{fig:5}
\end{center}
\end{figure}
\clearpage

\begin{figure}
\begin{center}
\FigureFile(160mm,80mm){figure6.eps}
\caption{(a) Distributions of massive stars and young stellar objects (*1 \cite{1960MNRAS.121..337F}, *2 \cite{1970ApJ...161L.149W}, *3 \cite{1981MNRAS.197P..17G}, *4 \cite{1992A&A...259..480D}, *5 \cite{1995PASP..107..145C}, *6 \cite{2002MNRAS.331..969L}, *7 \cite{2005ApJ...620..731J}, *8 \cite{2005A&A...433..205M}, *9 \cite{2005AJ....129..776N}, *10 \cite{2007A&A...469..459T}, *11 \cite{2006A&A...453..517T} overlaid with H$\alpha$ image. (b) Distributions of cluster with nebulosity cataloged by \citet{1999AJ....117..238B}.
}
\label{fig:6}
\end{center}
\end{figure}
\clearpage

\begin{figure}
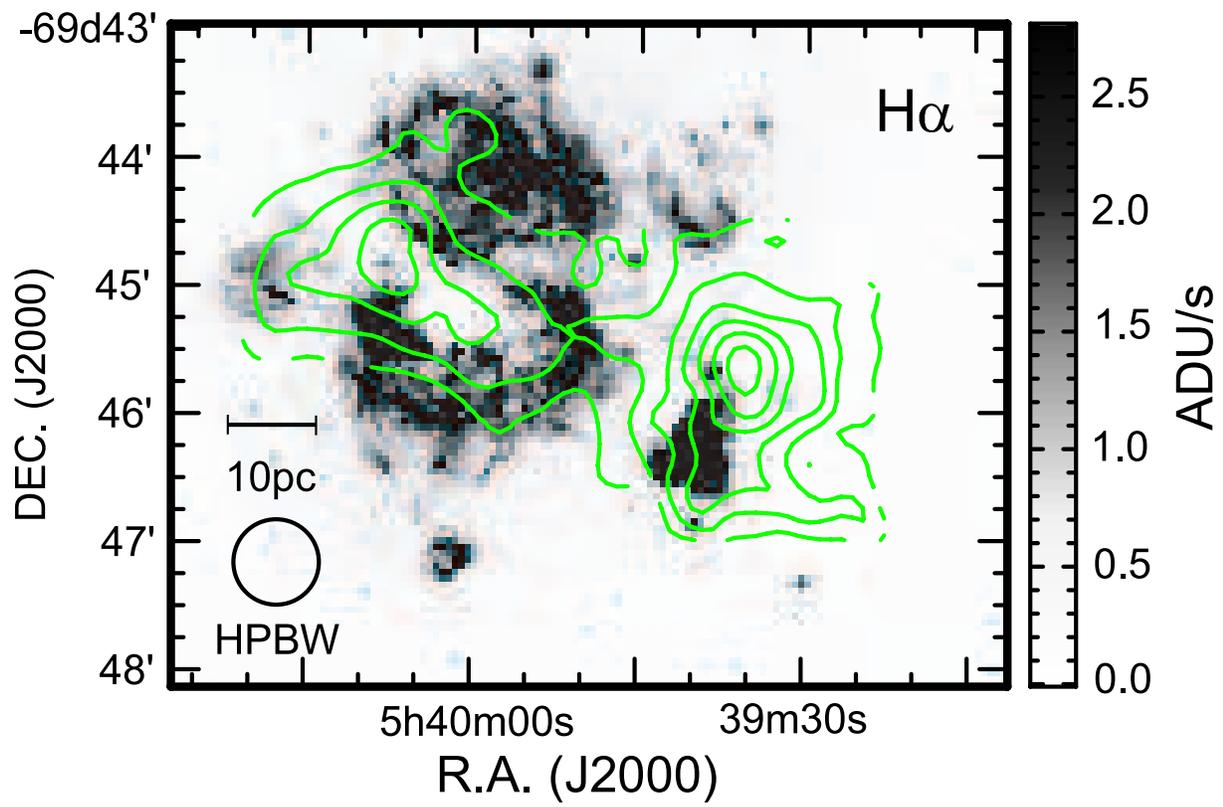

\begin{center}
\FigureFile(160mm,80mm){figure7.eps}
\caption{$^{12}$CO $J$=4--3 integrated intensity map overlaid with H$\alpha$ image.}
\label{fig:7}
\end{center}
\end{figure}
\clearpage

\begin{figure}
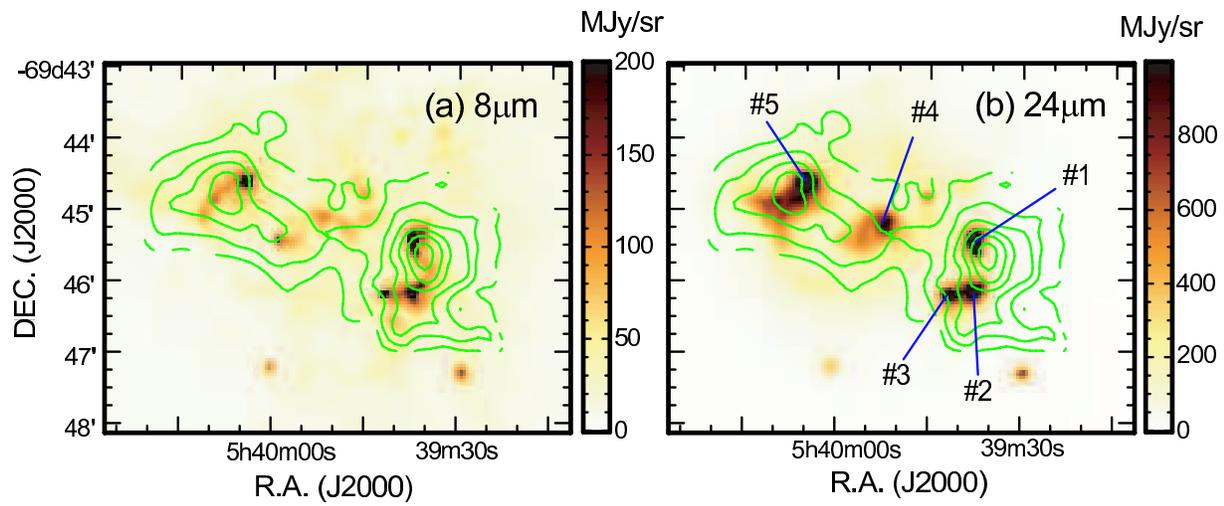

\begin{center}
\FigureFile(160mm,80mm){figure8.eps}
\caption{Contour of $^{12}$CO $J$=4--3 integrated intensity overlaid with 8$\micron$ (a) and 24$\micron$ (b) image from Spitzer/SAGE. Properties of marked peak are listed in Table \ref{tbl:7}.}
\label{fig:8}
\end{center}
\end{figure}
\clearpage

\begin{figure}
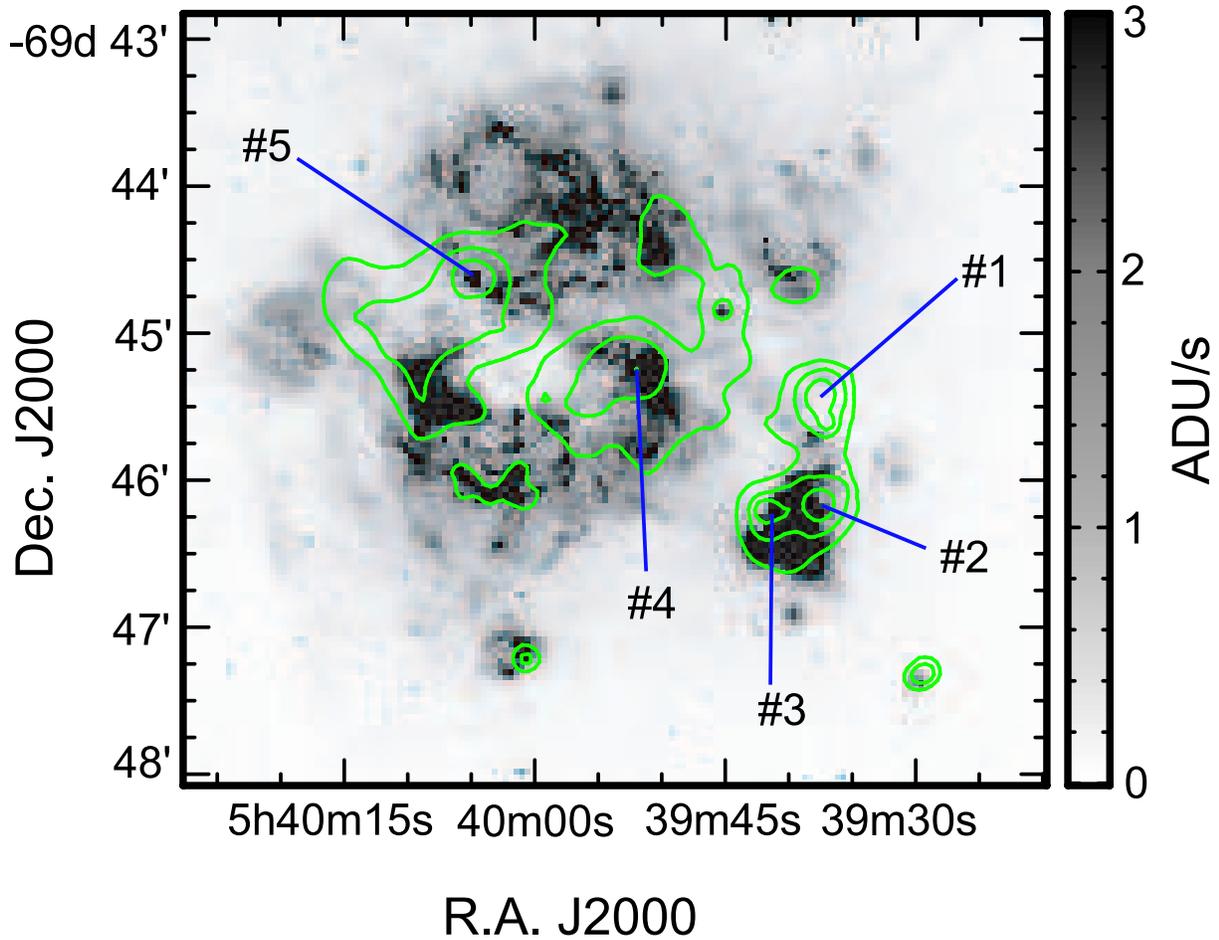

\begin{center}
\FigureFile(160mm,80mm){figure9.eps}
\caption{Diagram shows H$\alpha$ image and contour of 24$\micron$ flux. Contour levels are 200, 400, 800, 1600 MJy/sr.}
\label{fig:9}
\end{center}
\end{figure}
\clearpage

\begin{figure}
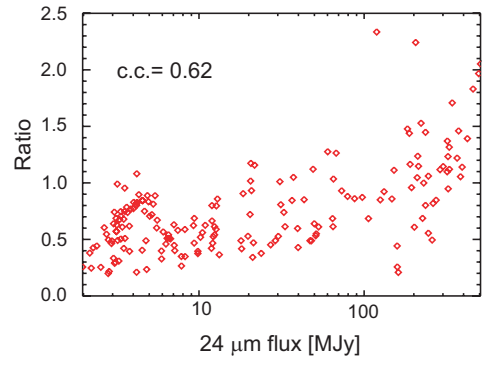

\begin{center}
\FigureFile(80mm,80mm){figure10.eps}
\caption{Pixel by pixel comparison (20$\arcsec$ grid) of line ratio R$_{3-2/1-0}$ with 24 $\micron$ flux, where the correlation coefficient is 0.62.}
\label{fig:10}
\end{center}
\end{figure}
\clearpage

\begin{figure}
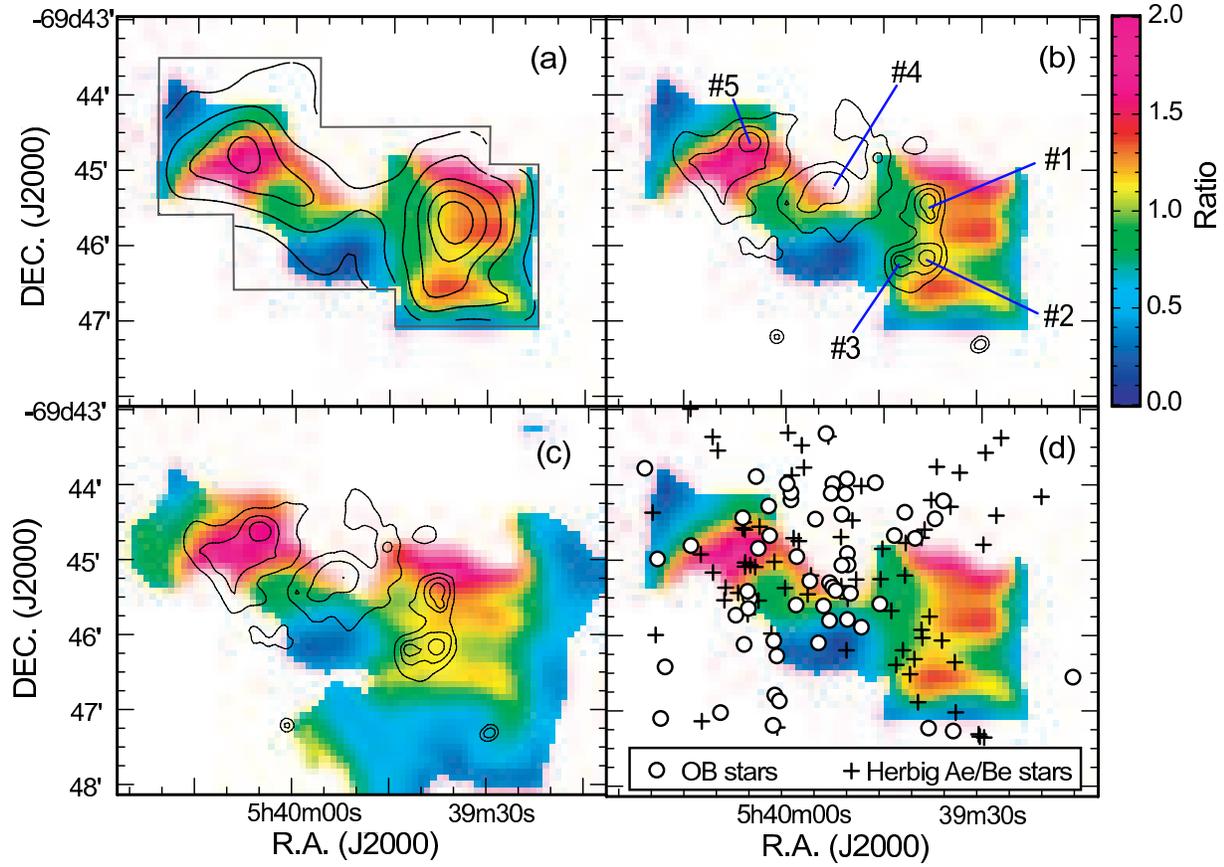

\begin{center}
\FigureFile(160mm,80mm){figure11.eps}
\caption{(a) Image of integrated intensity ratio R$_{4-3/1-0}$ and contour of $^{12}$CO $J$=4--3 integrated intensity in northern part. (b) Contour of 24 $\micron$ flux superposed on R$_{4-3/1-0}$ and marked number show 24$\micron$ intensity peak position. (c) R$_{3-2/1-0}$ and contour of 24 $\micron$ flux. (d)  Positions of OB stars and Herbig Ae/Be stars derived by \citet{2005AJ....129..776N} overlaid with R$_{4-3/1-0}$.}
\label{fig:11}
\end{center}
\end{figure}
\clearpage

\begin{figure}
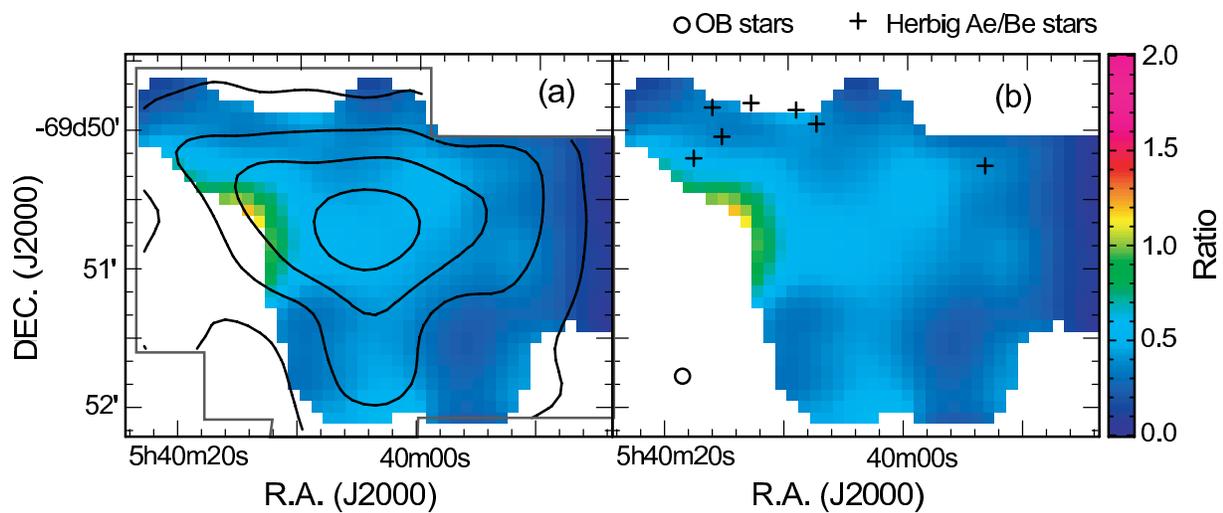

\begin{center}
\FigureFile(160mm,80mm){figure12.eps}
\caption{Same as Figure \ref{fig:11} but in southern part.}
\label{fig:12}
\end{center}
\end{figure}
\clearpage

\begin{figure}
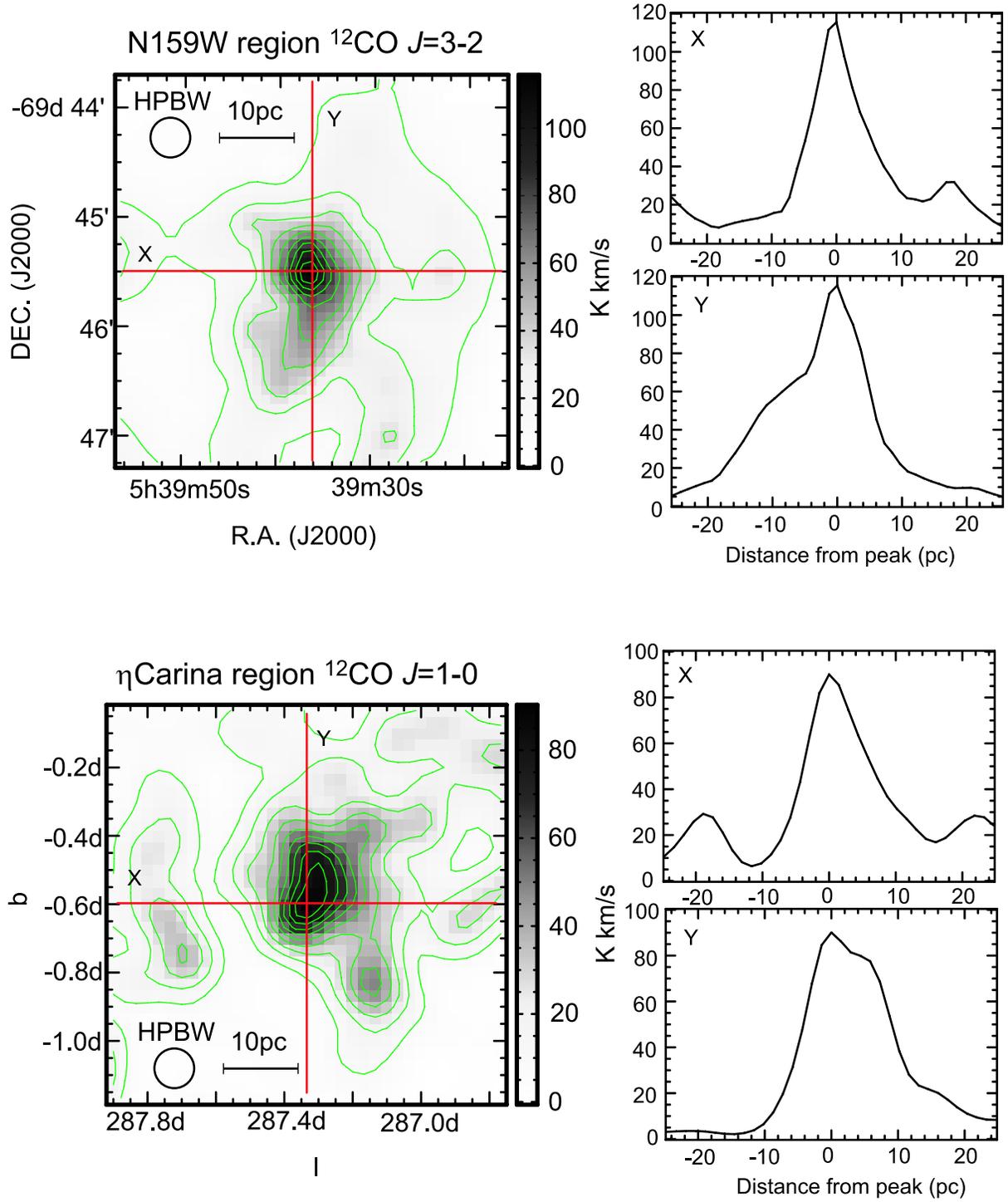

\begin{center}
\FigureFile(160mm,80mm){figure13.eps}
\caption{Molecular gas distributions of N159 (top) and $\eta$ Car, northern cloud \citep{2005ApJ...634..476Y}. The map of $\eta$ Carinae region was smoothed to same spacial resolution of N159 (5pc). Contour levels are 10\%, 20\%, ..., 90\% of peak integrated intensity. Intensity distribution in the directions of X, Y slit in the left figure are displayed in right figure.}
\label{fig:13}
\end{center}
\end{figure}
\clearpage

\end{document}